\documentclass{aa}  

\usepackage{graphicx}
\usepackage{txfonts}
\usepackage{multirow}
\usepackage{lscape}
\usepackage{natbib}
\bibpunct{(}{)}{;}{a}{}{,} 
\usepackage[dvipsnames]{color}

\begin{document}

   \title{Ray-tracing GR-MHD-generated outflows from AGNs hosting thin accretion disks}

   \subtitle{An analysis approaching horizon scales}

   \author{Bidisha Bandyopadhyay\inst{\ref{1}}\thanks{bidisharia@gmail.com (BB)}\and 
          Christian Fendt\inst{\ref{2}}\thanks{fendt@mpia.de (CF)} \and
          Dominik R.G. Schleicher\inst{\ref{3},\ref{1}}\and
          Neil M. Nagar\inst{\ref{1}}\and
          Felipe Agurto-Sepúlveda\inst{\ref{4}}\and
          Javier Pedreros\inst{\ref{1}}
          }

   \institute{Departamento de Astronom\'ia, Facultad Ciencias F\'isicas y Matem\'aticas, Universidad de Concepci\'on, Av. Esteban Iturra s/n Barrio Universitario, Casilla 160-C, Concepci\'on, Chile \label{1}            
         \and
             Max Planck Institute for Astronomy, K\"onigstuhl 17, D-69117 Heidelberg, Germany,\label{2}
            \and
        Dipartimento di Fisica, Sapienza Università di Roma, Piazzale Aldo Moro 5, 00185 Rome, Italy\label{3}
            \and 
            Departamento de Físca de Partículas y Instituto Galego de Física de Altas Enerxías (IGFAE), Universidade de Santiago de Compostela, E-15782 Santiago de Compostela, Spain \label{4}}

   \date{}

  \abstract
   {Active galactic nuclei (AGNs) exhibit a wide range of black hole masses and inflow/outflow properties. It is now possible to probe regions close to the event horizons of nearby supermassive black holes (SMBHs) using very long baseline interferometry (VLBI) with earth-sized baselines, as performed by the Event Horizon Telescope (EHT).}
   {This study explores the emission properties of accretion and outflows near the event horizon of both low-mass and high-mass SMBHs. Using resistive general relativistic magnetohydrodynamic (GR-MHD) simulations, we model AGNs with thin Keplerian disks. This contrasts with widely studied models featuring thick disks, such as magnetically arrested disks (MADs) or the standard and normal evolution (SANE) scenario.}
   {Our GR-MHD models serve as simplified representations to study disk–jet–wind structures. These simulations are postprocessed and ray-traced, using constraints of black hole mass and observed spectral energy distributions (SEDs). Thermal synchrotron emission generated near the event horizon is used to create emission maps, which are analysed by separating accretion and outflow components to determine their contributions to the total intensity.
    }
   {Whether the emission appears optically thick or thin at a given frequency depends on its position relative to the synchrotron SED peak. At 230 GHz, low-mass SMBHs appear optically thicker than high-mass ones, even at lower accretion rates. Doppler beaming affects the brightness of emission from outflows with changing viewing angles in low-mass systems.}
   {Eddington ratios from our models align with those inferred by the EHTC for M87 and SgrA* using thicker MAD/SANE models. Although thin disks are optically thicker, their spectral properties make high-mass systems appear optically thinner at 230 GHz—ideal for probing GR effects like photon rings. In contrast, low-mass systems remain optically thicker at these frequencies because of synchrotron self-absorption, making outflow emissions near the horizon more pronounced. However, distinguishing these features remains challenging with current EHT resolution.}

   \keywords{Galaxies: nuclei --
                Black hole physics --
                Galaxies: jets
               }

   \titlerunning{Ray-tracing AGN outflows on Horizon scales}
   \authorrunning{B. Bandyopadhyay et al.}
   \maketitle
   
%

\section{Introduction}

Astrophysical black holes (BHs) are found with masses ranging from a few solar masses to supermassive BHs (SMBHs) of masses greater than $10^{6} M_{\sun}$. They have been predicted on the basis of Einstein's general theory of relativity (GR) defined via a one-way causal boundary in spacetime. This is also known as the event horizon from which even light cannot escape, making the direct detection of BHs impossible. 
It has been found that BHs can be detected only indirectly by their gravitational influence on the surrounding matter and/or spacetime, by accretion that results in often luminous emission from radio to $\gamma$-rays,  or by gravitational-wave emission.

Event Horizon Telescope (EHT) has delivered the first images of the gravitationally lensed ring and the BH shadow of the SMBH of M87 \citep{EHTC2019a,EHTC2019b,EHTC2019c,EHTC2019d,EHTC2019e} in the heart of the Virgo cluster and of SgrA* at the center of our Galaxy \citep{,EHTC2019f,EHTC2022a,EHTC2022b,EHTC2022c,EHTC2022d,EHTC2022e,EHTC2022f}.
Active galactic nuclei (AGNs), which are accretion-powered SMBHs in the nuclei of some galaxies, have intrigued astronomers for ages as the most powerful sources in the sky. 
They are able to outshine even the entire stellar population of the galaxy that hosts them, such as quasars, which are among the most luminous sources in the sky \citep{Schmidt1963, Sanders1989, Novikov1973}.
These are believed to be powered by high accretion rates onto the SMBH through a geometrically thin but optically thick disk \citep{Shakura1973, Sun1989}. On the other hand, AGNs in the Local Universe, including M87 and SgrA*, predominantly show lower accretion rates onto the central BH \citep{Ichimaru1977, Narayan1995, Blandford1999, Nagar2005, Yuan2014}. 

Many AGNs display signatures of outflows and large-scale jets, which can be observed throughout the spectrum. 
Jets appear as collimated structures launched at relativistic velocities from very close to the BH. 
 EHT recently imaged the innermost jets of 3C 279, CenA, and the blazar J1924-2914 \citep{EHTC2020,Janssen2021,Issaoun2022} with a 20 $\mu \rm as$ resolution, paving the way to improve our observation and understanding of the physical processes involved in launching jets around SMBHs. 
Jets may either be powered by magnetic fields that are threaded throughout the event horizon, by extracting the rotational energy of the BH \citep{Blandford1977}, or by the magneto-rotational acceleration of matter from the accretion flow \citep{Blandford1982}. 
The presence of strong magnetic fields close to the event horizon thus leads to synchrotron radiation, which peaks between the radio and the far-infrared \citep{Ho1999} and could originate from the accretion flow, jet base, or disk wind.

The first-ever images of the BH shadows of two AGNs with contrasting masses, namely, M87 and SgrA*, have paved the way for understanding the physical processes around SMBHs. Although both sources have roughly the same flux and angular size in their gravitationally lensed rings, they are individually interesting because their masses and, potentially, their spins are on the two extreme ends of the mass and spin spectrum of SMBHs. 

Although the BH mass of M87 ($M_{\rm BH}=6.5 \times 10^{9}~M_{\odot}$) is three orders of magnitude higher than that of SgrA* ($M_{\rm BH}=4.16 \times 10^{6}~M_{\odot}$), it is also approximately three orders of magnitude farther away ($D_{M87}=16.8$ Mpc) than SgrA* ($D_{SgrA*}=8.127$\,kpc) [see, e.g., \citet{EHTC2019a, EHTC2022a}]. 
This gives rise to interesting observational similarities, such as their lensed photon rings project similar angular sizes on Earth 
($\simeq 42~\mu$arcsec for M87 from \citealt {EHTC2019f} and $\simeq 52~\mu$arcsec for SgrA* from \citealt{EHTC2022a}). These intriguing aspects place them among the key targets for EHT.

Although our telescopes capture similar fluxes for both these sources, M87 is intrinsically much more luminous compared to SgrA*. This could be attributed to the total accretion power of M87, which is estimated to be approximately $10-100$ times greater than SgrA* \citep{EHTC2021}. The extended one-sided M87 jet and the relative faintness of the nuclear counter-jet provide strong constraints on the orientation of the jet and, thus, on the spin axis of the black hole: an inclination of $\simeq 20\degr$ to the line of sight was used for numerical simulations of M87 \citep{EHTC2019e}. 
For SgrA*, there are  no strong constraints set on the spin axis yet, although the model fits of the EHTC suggest an inclination angle of less than $30\degr$ to the line of sight \citep{EHTC2022a}. These interesting characteristics of M87 and SgrA* make them the best candidates for testing different theoretical models.  

The Event Horizon Telescope, in science operation since 2017, combines a network of globally distributed mm-wave telescopes using the very large baseline interferometry (VLBI) technique. Currently 11 telescopes participate in the EHT, with a twelfth (AMT, Namibia) now funded \citep{labella2023}. Resolutions for VLBI can be significantly improved using low Earth orbit (LEO) satellites \citep{Roelofs2021}, Earth-Geostationary \citep[GEO-VLBI;][]{Fish2020} and/or Earth-L2 \citep[L2-VLBI;][]{Likhachev2022} baselines. 

In addition to improving image quality with extended baselines, the next-generation EHT (ngEHT) also plans to improve image construction algorithms from the data to be able to obtain the strongest possible constraints on the complex structures around the BH event horizon \citep[e.g.,][]{Roelofs2023}.
The extended baseline obtained with future Earth-space-based VLBI (a possibility through space-based ngEHT) will result in an enhancement in the resolution which will allow us to distinguish between various physical scenarios. 

Sources such as M87 and Sgr A* are classified as low-luminosity AGNs (LLAGNs), which are believed to be best described by advection-dominated accretion flow models (ADAF) \citep{Yuan2014}. In these cases, because of the low density, the ions and electrons are not in thermal equilibrium with each other, and excess heat is advected onto the BH. 
The MAD/SANE GR-MHD models \citep{Igumenshchev2003,Narayan2012}, mostly used to describe systems such as M87 or Sgr A*, are the closest to an ADAF-like scenario.
In that case, the accretion flow forms a torus-like structure, leading to an accretion flow that is geometrically thick but optically thin. 

In the present work, however, we investigate resistive GR-MHD models with thin Keplerian disks known to be able to produce strong outflows. 
Our foremost interest lies in investigating the emission close to the horizon scales in such scenarios, looking for the possibility of identifying emission signatures from outflows and possibly disentangling the radiation from the different systemic components. 

In this work, we intend to distinguish and compare the emission arising from a range of dynamical models with thin Keplerian disks for a high-mass and a low-mass BH system. To this extent, we therefore considered the BH masses and observed spectral energy distribution (SED) data of M87 and SgrA* \citep{Narayan1998, Prieto2016}] as the constraint parameters. We postprocessed the results from a set of GR-MHD models, thus scaling them to physically realistic systems to investigate the emission features generated. The systems studied here are thus referred to as HIGH-mass and LOW-mass systems.

Our paper is organized as follows. In Sect. \ref{subsec:dyn}, we describe the dynamical setup of our simulations. In Sect. \ref{subsec:raytr}, the postprocessing and ray-tracing methods are discussed, especially in the context of those chosen for this work. We also describe the process of separating the emissions from the accreting disk and the outflow. In Sect. \ref{sec:Results}, we discuss the primary results obtained by fitting the synthetic spectra to the observed data, as well as the emission properties for the high and low mass systems under consideration. In Sect. \ref{sec:Discussion}, we provide a detailed physical analysis of the different results obtained, while in Sect. \ref{sec:Conclu}, we provide the main highlights obtained from this investigation.

\section{Methodology}

\begin{table*} 
\begin{center}
\caption{Characteristic parameters of our simulation runs.}
\label{tab:para_dynamics}
\begin{tabular}{ccccccccl}
\hline
\hline
\noalign{\smallskip}
\multirow{2}{*}{$ID$} & \multicolumn{5}{c}{Simulation inputs} & \multicolumn{2}{c}{$\dot{m}$ from SED fits} & \multirow{2}{*}{$Comments$} \\
 & $\beta_0 $ & $\rho_{\rm flr}$ & $u_{\rm flr}$  & $a$ & $R_{\rm isco}$ & $\dot{m}_{\rm HIGH-mass}$ &  $\dot{m}_{\rm LOW-mass}$ & \\
\noalign{\smallskip}
\hline
\noalign{\smallskip}
\noalign{\smallskip}
 SIM20  & 10  & $10^{-5}$ & $10^{-8}$  & 0.9375 & $2~R_{\rm g}$ & $2.0 \times 10^{-5}$ & $5.6 \times 10^{-7}$ & reference run, similar to \citet{Vourellis2019} \\
 SIM21  & 10  & $10^{-4}$ & $10^{-6}$  & 0.9375 & $2~R_{\rm g}$ & $1.0 \times 10^{-5}$ & $5.0 \times 10^{-7}$ & as SIM20, floor density higher \\
 SIM22  & 10  & $10^{-4}$ & $10^{-6}$  & 0.0    & $6~R_{\rm g}$ & $7.0 \times 10^{-5}$ & $5.0 \times 10^{-6}$ & as SIM20, $a=0$, infall to BH, disk wind \\
 SIM23  & 10  & $10^{-3}$ & $10^{-5}$  & 0.9375 & $2~R_{\rm g}$ & $3.0 \times 10^{-6}$ & $1.5 \times 10^{-7}$ & as SIM21, floor even higher, no BZ \\
 SIM24  &  1  & $10^{-3}$ & $10^{-5}$  & 0.9375 & $2~R_{\rm g}$ & $4.0 \times 10^{-7}$ & $2.0 \times 10^{-8}$ & as SIM21, $\beta_0$ lower, stronger magn. field \\
 SIM26  & 0.1 & $10^{-3}$ & $10^{-5}$  & 0.9375 & $2~R_{\rm g}$ & $2.0 \times 10^{-7}$ & $1.0 \times 10^{-8}$ & as SIM24, $\beta_0$ even lower, even stronger magn. field \\
  \noalign{\smallskip}
 \hline
 \noalign{\smallskip}
 \end{tabular}
\end{center}
\tablefoot{Shown here: the simulation ID, the initial disk plasma beta, $\beta_0$ (at the inner disk radius),
the maximum floor density, $\rho_{\rm flr}$, and internal energy, $u_{\rm flr}$, 
the Kerr parameter of the BH, $a$, the radius of the ISCO, $R_{\rm isco}$, and specific comments on the simulation runs, respectively. 
The Eddington ratios of these simulations for the defined HIGH-mass and LOW-mass system 
are obtained from the SED fits to the data for M87 as tabulated in \citet{Prieto2016} and for SgrA* in \citet{Narayan1998}. The fits are obtained by fixing the values of $R_{low}=1$, $R_{high}=80$ and an inclination angle of $i=17^{\circ}$ when postprocessing the simulated data for each of the models.}
\end{table*}

Our approach follows a two-step procedure. 
First, we carried out a set of resistive general relativistic magnetohydrodynamic (GR-MHD) simulations of the disk-jet evolution for a range of parameters. 
Then we postprocess the final snapshots of the dynamical simulation data applying ray-tracing procedures. 
We note that the dynamical data are scale-free by definition, as these are pure magnetohydrodynamic simulations.
We thus needed to rescale our physical parameters, such as the mass density, gas pressure, and magnetic field strength, to astrophysical units.
As a result, this approach allowed us to apply our dynamical models to accreting BH systems with various physical conditions, such as BHs of different mass, accretion rates, and magnetic flux. 
In turn, we were able to derive these parameters from the BHs by comparing our results to the observed systems.

\subsection{Dynamical modeling}\label{subsec:dyn}
To provide a fiducial set for the magnetohydrodynamic structure of accreting systems, we performed a set of GR-MHD simulations
for the accretion-ejection system surrounding a central black hole.
We choose a setup that should be understood as a "toy model" for the very center of the AGN, 
as it is able to generate clearly structured components, including: the black hole, the (thin) accretion disk, a strong disk wind, and a central spine jet.
The model allows to govern the appearance of these components by choosing central physical parameters such as the magnetic field strength or the mass flux of the spine jet.

We applied the resistive GR-MHD code rHARM3D
\citep{Qian2017, Qian2018, Vourellis2019}, 
which is an extension of the well-known HARM code \citep{Gammie2003, McKinney2004, Noble2006,Noble2009}.
When treating the transition from accretion to ejection of the disk material, 
the application of a physical resistivity in the GR-MHD equations is essential, as it allows for both a long-term mass loading of the disk wind and jet and a smooth accretion process. 
The resistivity is connected to the magnetic diffusivity.  
In comparison to ideal MHD, in resistive MHD the mass flux can diffuse across the magnetic field and can be diverted from accretion to ejection \citep{Ferreira1997}.
The mass flux that is ejected as a disk wind can thus be replenished by the mass that is advected by disk accretion.
This concept of resistive accretion-ejection structures is well investigated and widely applied for non-relativistic simulations
of jet launching \citep{Keppens2002, Zanni2007, Sheikhnezami2012, Stepanovs2016}
and has only recently been applied to GR-MHD \citep{Qian2017, Qian2018, Ripperda2019}. 
The interrelation between disk accretion and ejection rates and the magnitude of the applied resistivity has been investigated in detail in the aforementioned papers. We note that in addition to the redistribution of mass flux across the magnetic field, magnetic diffusivity allows us to treat reconnection in a way that is physically well defined.

Compared to the literature on GR-MHD accretion that mostly considers the well-known SANE (standard and normal evolution) or MAD (magnetically arrested disk, \citealt{Narayan2003}),
our approach is different in applying a {"}resistive thin Keplerian{"} disk.
The advantage of applying a thin disk as an initial condition for the simulation is that it allows to launch and accelerate strong disk winds \citep{Blandford1982, Ferreira1997,Keppens2002, Qian2018, Dihingia2021}.

We emphasize again that our intention is to model the emission features of these structural components and to compare them.
We note that it is not our intention here to fit the observed features of specific sources using a thin-disk model.

As our simulation results show (see also \citealt{Bandyopadhyay2021}), the disk accretion-ejection structure reaches a dynamical equilibrium state that allows for continuous disk accretion and wind ejection. 
Details of this equilibrium state naturally depend on the parameters of disk physics such as magnetic resistivity $\eta$ or plasma-$\beta$.
The GR-MHD models that we used in this study are the same as in \citet{Bandyopadhyay2021}
(see Table~\ref{tab:para_dynamics}).
Magnetic diffusivity or resistivity is understood as anomalous and thus to be caused by disk turbulence, similar to the $alpha$-viscosity of \citet{Shakura1973}.
A background diffusivity of $\eta_{\rm back}=10^{-3}$ was applied, as well as a maximum disk magnetic diffusivity of $\eta_{\rm max}=0.01$. 
We generally applied a density contrast of $\rho_{\rm cor} / \rho_{\rm disk} = 0.001$ at the inner radius of the disk.

The GR-MHD simulations considered here are axisymmetric.
Thus, the three spatial vector components are considered, but derivatives in the $\phi$ direction are neglected.
We note that resistive MHD simulations are exceptionally expensive for the CPU.
However, we believe that our approach is fully sufficient for our aim to investigate the inner jet launching area of a black hole-disk-wind-jet system applying GR radiative transport calculations and to derive radiative features from these system components.
Of course, any 3D features like disk or jet instabilities cannot be treated, nor any orbiting substructures can be produced.

It is worth summarizing the differences of our dynamical model setup in comparison to other attempts to model intensity maps and spectra
from GR-MHD simulations, such as those published in \citet{Dexter2012, Moscibrodzka2016, EHTC2019e}.
We work with resistive GR-MHD, which allows for smooth disk accretion and the launch of a disk wind.
Our disk is geometrically thin, in Keplerian rotation, and is threaded by a large-scale magnetic flux.
This allows for driving strong disk winds (which, on larger spatial scales, are expected to evolve into a jet). 
In these cases, it is the magnetized disk wind that is more efficient in angular momentum removal due to its large lever arm.

Unlike the standard hot accretion flows, where the inflow plasma is optically thin through out, in the model setup presented here, the density distribution of the gas is such that the density maximum lies around the disk plane; it then decreases in the direction away from the disk plane, with the density dropping by almost two orders of magnitude at $\simeq~20~R_g$ (see the density plots in \citealt{Bandyopadhyay2021}). 
Thus, these models are optically thicker, and as will be seen later the disk luminosity in some cases is higher (depending on the postprocessing parameters used) in contrast to optically thin hot-accretion flow models.

A common feature of all GR-MHD simulations is the necessity of applying a so-called floor density that is set as a lower density threshold in order to keep the MHD approach valid and the simulation going.
This numerical artifact may somewhat affect the resulting dynamics of the axial Blandford-Znajek (BZ) jet, 
that is, the mass load carried away with that jet and, consequently, also the jet velocity \citep{Qian2018, Vourellis2019}. 
However, the dynamics of the disk wind is governed by the mass loading from the disk (and the magnetic field initially assumed),
which is self-consistently derived from resistive MHD.

Our simulations follow the setup of \citet{Vourellis2019}.
The output of our GR-MHD simulations is in normalized units, where $G=c=M=1$.
In particular, all length scales used in these simulations are in gravitational radii $R_{\rm g}$, while velocities are normalized to the speed of light.
All physical variables must be properly rescaled during postprocessing to account for the radiation.

A further discussion of our model approach that includes the parameter space involved, as well as maps of the essential dynamical variables, can be found in \citet{Bandyopadhyay2021}.
To summarize our models, we investigate
(i) a reference simulation labeled as SIM20 which applies similar simulation parameters as \citet{Vourellis2019};
(ii) a simulation SIM21 with a (ten times) higher floor density, thus a higher mass loading of the BZ jet; 
(iii) a simulation SIM22 with a Kerr parameter $a=0$;
(iv) a simulation SIM23 with an even higher floor density (another factor ten) resulting in the absence of BZ jet; 
(v) a simulation SIM24 with a lower plasma-$\beta$ (factor ten), resulting in a stronger magnetic flux; and 
(vi) a simulation SIM26 with an even stronger magnetic flux (another factor of 10),
as shown in Table \ref{tab:para_dynamics}.

The model setups we applied mainly differ in terms of the magnetic field strength,  black hole spin, and the floor density.
In fact, these parameters have an immediate impact on the jet dynamics, namely: the jet speed (Doppler boosting) and mass flux; on the existence of a two-component jet (black hole spin). 
The synchrotron emission is regulated by the astrophysical scaling of the simulation variables, such as the physical density and the physical magnetic field strength.

\begin{figure*}
 \begin{center}

 \includegraphics[height=3.2in]{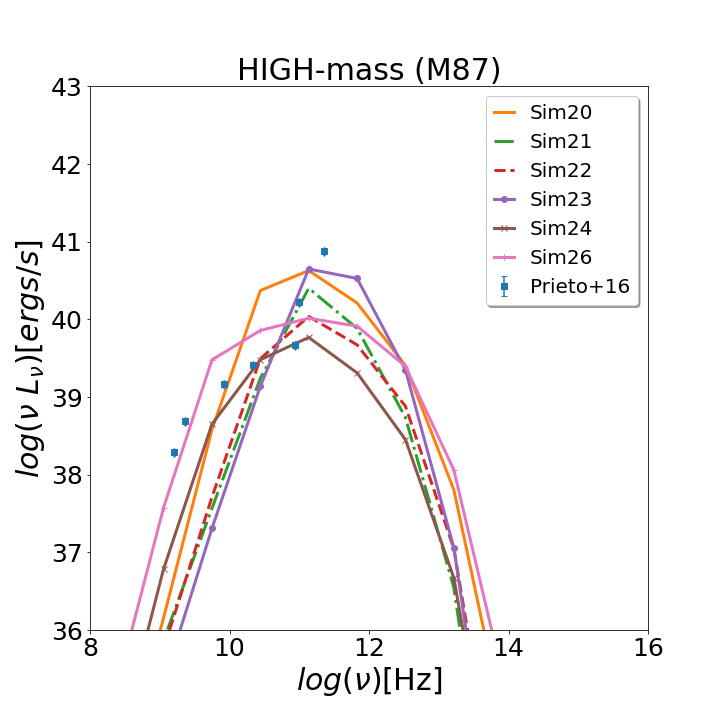}
  \includegraphics[height=3.2in]{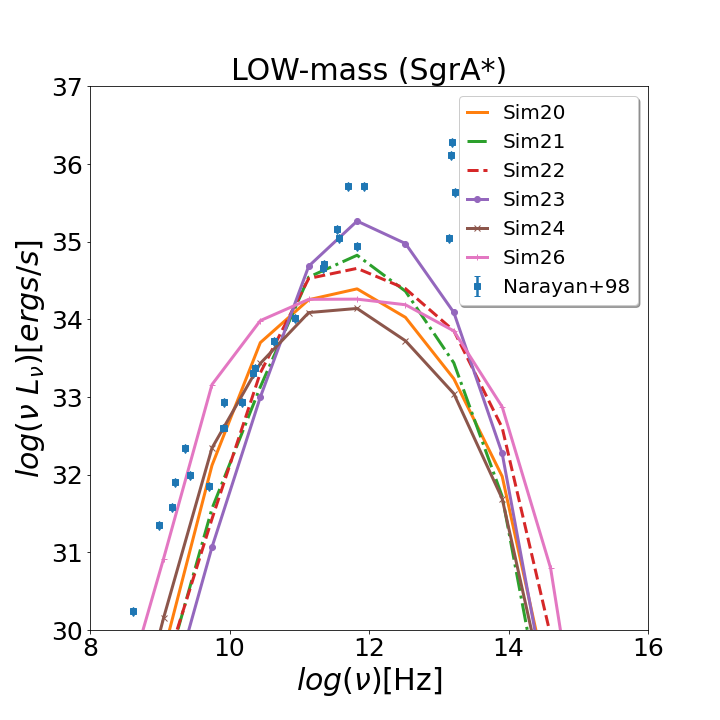}
  \caption{Thermal synchrotron spectra for the simulated dynamical models obtained for HIGH-mass (left) and LOW-mass (right) system parameters.  
  The Eddington ratio obtained for each of the simulations by fitting the model spectra to the data for HIGH-mass \citep{Prieto2016} and LOW-mass systems \citep{Narayan1998} is shown in the columns 7 and 8 of Table \ref{tab:para_dynamics}. These fits are obtained by fixing the postprocessing parameters $R_{low}=1$, $R_{high}=80$ and an inclination angle of $i=17^{\circ}$.}
  \label{fig:spectrathM87SgrA}
\end{center}
\end{figure*}

\subsection{Postprocessing, parameter fitting, and ray-tracing}\label{subsec:raytr}
GR-MHD codes typically apply normalized units where $G = c = M = 1$. 
Thus, these simulations need to be postprocessed with astrophysical parameters such as black hole mass, accretion rates, and so on -- to obtain a physical realization of an AGN. 
Here we postprocess the final snapshots of the simulations (i.e., when the simulation attains a quasi-steady configuration) 
for BH masses as for M87 or SgrA* and vary the accretion rate (that governs other physical parameters) to obtain synthetic SEDs.
These are compared with the observed data in the radio bands as the emission originating from regions in the vicinity of the BH is expected to be highly magnetized, and thus it is also dominated by synchrotron processes as a result \citep{Yuan2014}. 
The postprocessing of these GR-MHD simulations is performed using the GRTRANS ray tracing code \citep{Dexter2016}.
 
The image of a BH is characterized by a {"}shadow{"} region from which no photons arrive due to the curvature of spacetime around the BH and the unlensed "photon ring" ($3~R_g$), corresponding to bound photon orbits around the BH. As a result of the lensing, the photons from the inner part of the disk appear somewhat further outside that region, leading to a "lensed ring" marked by inner ($5.2~R_g$) and outer ($6.2~R_g$) boundaries \citep{Gralla2020a, Gralla2020b, Gralla2020c, Gralla2019, Johannsen2013, Beckwith2005, Luminet1979, Bardeen1973}. These features are a result of strong GR effects around the vicinity of the BH. 

The use of a ray-tracing code is to postprocess the simulated dynamical model to calculate the observed intensity in locations (pixels) of an observer’s camera for a given emission and absorption model. Here, we use the publicly available general relativistic radiative transport code GRTRANS \citep{Dexter2016}. Within GRTRANS, the Boyer–Lindquist coordinates of the photon trajectories are calculated from the observer towards the BH (i.e., the rays are traced) for each pixel in the camera. The observed polarization basis is then parallel-transported into the fluid frame. The local emission and absorption properties are calculated at each point. 
The radiative transfer equations are solved along those rays. We refer to the original paper of \citet{Dexter2016} for details on the working of the code.

Synchrotron emission can arise from different populations of electron distribution functions. However, a synchrotron emission arising from a population of thermal electrons as the primary emission process exists throughout the magnetized regions. This emission, often referred to as thermal synchrotron emission, depends on the thermal state of the electrons, where the electron temperature should ideally be determined from plasma physics. However, setting up simulations of a fully functional plasma dynamics in a GR setup is computationally expensive. In the absence of such a physically complete model, an alternative prescription is typically applied, which replicates the plasma temperature model. In this scenario, the gas temperature at each point in the grid is determined from the internal energy at those grid points assuming an ideal gas prescription. The electron temperature is deduced from the gas temperature assuming that the plasma consists only of ionized hydrogen. As prescribed by \citet{Moscibrodzka2009,Moscibrodzka2016}, the electron temperature is approximated by the following relation:

\begin{eqnarray}
    T_{e}&=&T_{\rm gas}/(1+R), \nonumber\\
    R & = & \frac{T_{\rm p}}{T_{\rm e}} = R_{\rm high}\frac{b^2}{1+b^2}+R_{\rm low}\frac{1}{1+b^2},
\label{eqn:temp}
\end{eqnarray}
where $T_p$ is the ion temperature, $b=\beta/\beta_{\rm crit}$, $\beta=P_{\rm gas}/P_{\rm mag}$ and $P_{\rm mag}=B^2/2$. The value of $\beta_{\rm crit}$ is assumed to be unity, and $R_{\rm high}$ and $R_{\rm low}$ are the temperature ratios that describe the electron-to-proton coupling in the weakly magnetized (e.g., disks) and strongly magnetized regions (e.g., jets), respectively.  

All GR-MHD simulations include floor density and pressure to avoid extreme low densities. However, this non-physical floor value leaves an imprint on the emission model. The magnetization parameter $\sigma_i$ is a ratio of the magnetic energy to the rest mass energy of the fluid as: $\sigma_i=|B|^2/4\pi \rho c^2$. Here, it has a value of $\sigma_i > 1$ in general in the accretion region close to the BH. The value of $\sigma_i$ exceeds the value of 1 in the jet (highly magnetized funnel region) close to the poles, which implies that the plasma
dynamics in those regions will be determined by the magnetic field. Thus, any small numerical glitch in the conserved energy in the simulation can lead to large errors in the energy of the fluid, affecting the plasma temperature as well. To avoid the unphysical impact of the floor values and the error resulting from the large magnetic fields, regions with $\sigma_i > 1$ are avoided \citep{Chael2018}.

As shown in \citet{Bandyopadhyay2021}, the postprocessing parameters that significantly affect the synthetic SED and emission maps are the BH mass and the accretion rate. The density, pressure, and magnetic field from the simulations are scaled according to the accretion rate and black hole mass. The temperature of the gas is obtained from the density and pressure assuming an adiabatic gas. The temperature and the magnetic field are the important physical parameters that affect the synchrotron emission through the distribution function (refer to \citealt{Dexter2016}). The SED was obtained for these models, covering regions that extend up to $100~R_g$ from the center. The reference data at different frequencies may have been obtained with a resolution and field of view (FOV) greater than or less than those assumed to obtain the total SED in this work, but for a broad comparison between the different models, an approximate fit to the data points at low frequencies is assumed to be correct. 

As shown in \citet{Bandyopadhyay2021}, the variation in $R_{low}$ affects the overall spectrum more than the variation in $R_{high}$, where lower values of $R_{low}$ or $R_{high}$ produce maximum flux. The usual approach adopted to obtain a perfect fit to the observed data is to choose a range of $R_{low}$ and $R_{high}$. Fast-accreting matter close to a SMBH is expected to be in an ionized state with the ions not in thermal equilibrium with the electrons \citep{Rees1982, Yuan2014}. The choice of $R_{low}=R_{high}=1$ assumes that electrons and ions have the same temperature. In this work, we assume $R_{low}=1$ and $R_{high}=80$ to avoid any significant changes in the SED as well as to observe the maximum features in the emission maps. Thus, keeping the values of $R_{high}$ and $R_{low}$ constant throughout the simulations, we varied the Eddington ratio in each case to obtain a reasonable fit to the observed SED for M87 \citep{Prieto2016} and SgrA* \citep{Narayan1998}. We used the Eddington ratios obtained after the spectral fit to further investigate the emission properties for a HIGH-mass and a LOW-mass system from these simulations. 

During postprocessing, we cut out any regions that were 10 degrees above and below the equatorial planes to separate out the contribution of the accretion disk from the "outflow." This is inspired by the procedure followed in \citet{Vourellis2019} as our simulation configurations are similar to those described in that work. Any region above or below this is defined as the outflow-dominated region (i.e., the non-accreting region). The contribution to the total emission from the accreting region is obtained by postprocessing and ray-tracing the GR-MHD data within this region. The contribution from the "outflow" region can then be obtained by removing the flux obtained from this region from the total flux. The resulting emission maps are shown in Fig. \ref{fig:allsimallidiskoutflow}. To obtain a more quantitative understanding of these emission regions, we obtained the radial emission profiles for several cross-sections, such as those along the X-axis and Y-axis and the two diagonals as shown in Fig. \ref{fig:prothdiskoutflowalli2} for an inclination angle of $17^{\circ}$.

In the next step, we explored whether the emission features from these models with outflows can be distinguished using current and future VLBI techniques. For this purpose, the synthetic ray-traced emission maps obtained were convolved with a Gaussian beam width to account for the different telescope resolutions. At 230 GHz, an intrinsic resolution (FWHM) of $\sim 20~\mu as$  was achieved for the longest baseline ($D_1\sim 11,000$ km). Thus, using these values, we were able to obtain approximate values of the resolution for Earth space baselines. A Geo-VLBI baseline ($D_2\sim47,000$~km) and a L2-VLBI baseline ($D_2\sim1.5$ million km) would give us approximate resolutions of $\theta_2\sim 5~ \mu as$ and $\theta_2\sim0.16~ \mu as$, respectively. We assumed these resolutions for the rest of this work, described in the sections below (see also \citealt{Pesce2021}).

\begin{figure*}
 \begin{center}
\includegraphics[height=2.35in,bb=10 10 650 650]{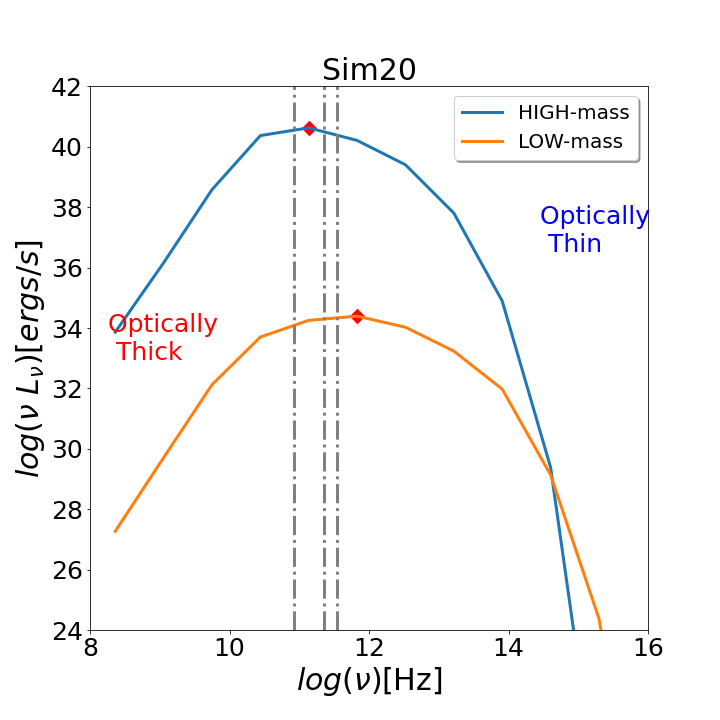}
\includegraphics[height=2.35in,bb=10 10 650 650]{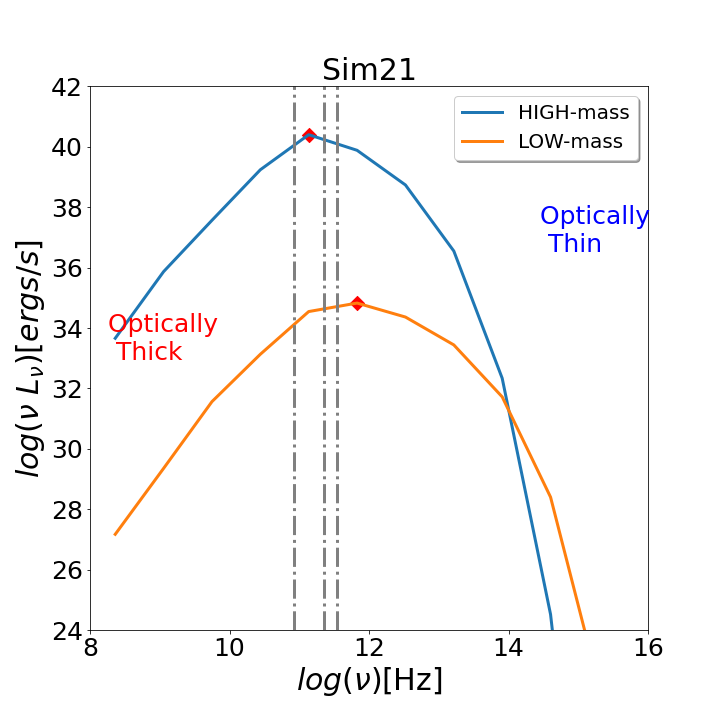}
\includegraphics[height=2.35in,bb=10 10 650 650]{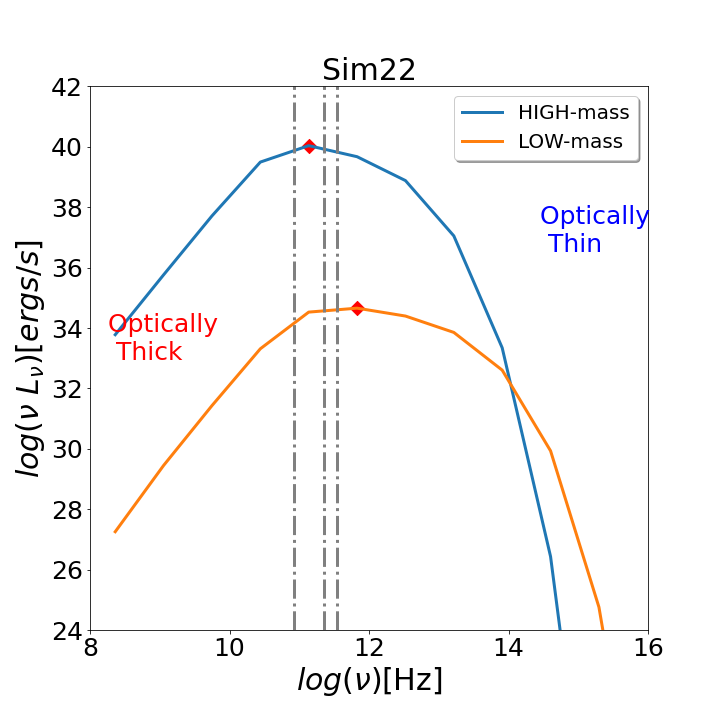} 
\vspace{0.07cm}
\\
\includegraphics[height=2.35in,bb=10 10 650 650]{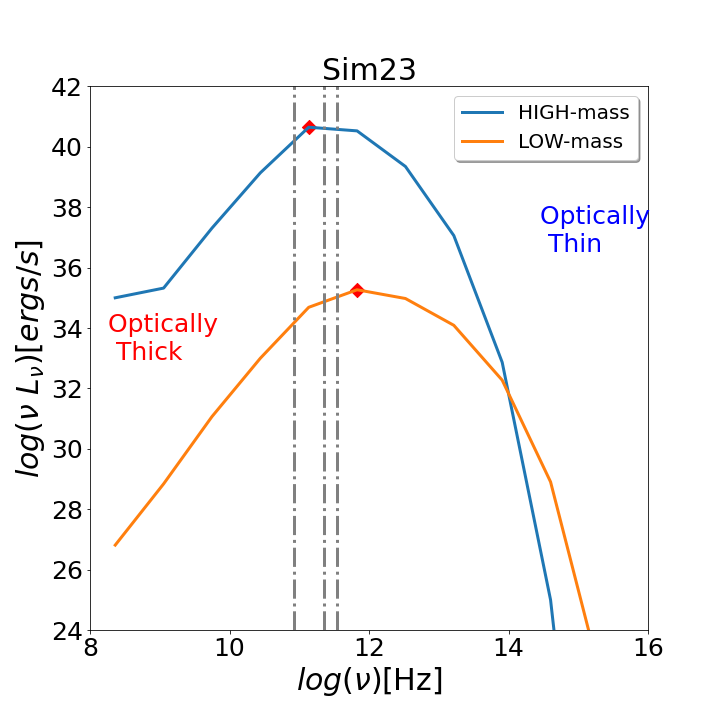} 
\includegraphics[height=2.35in,bb=10 10 650 650]{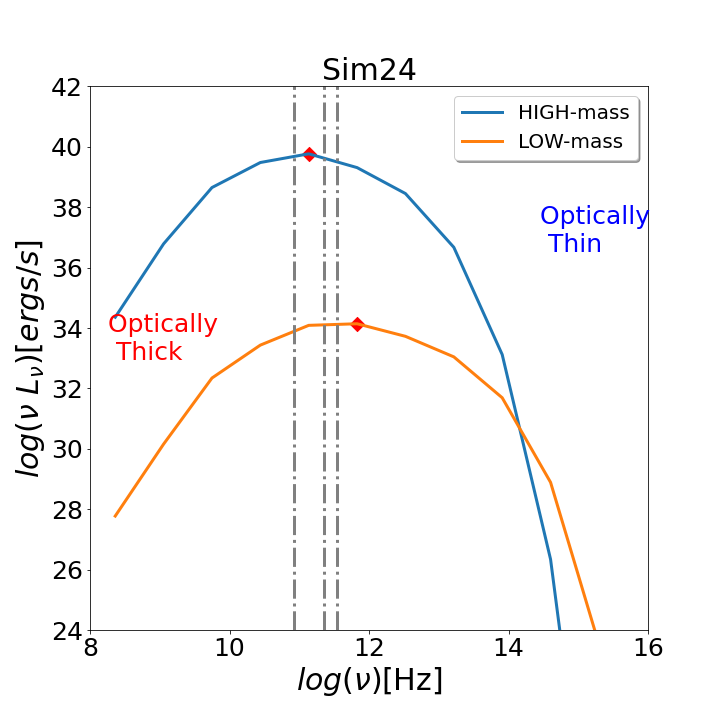}
\includegraphics[height=2.35in,bb=10 10 650 650]{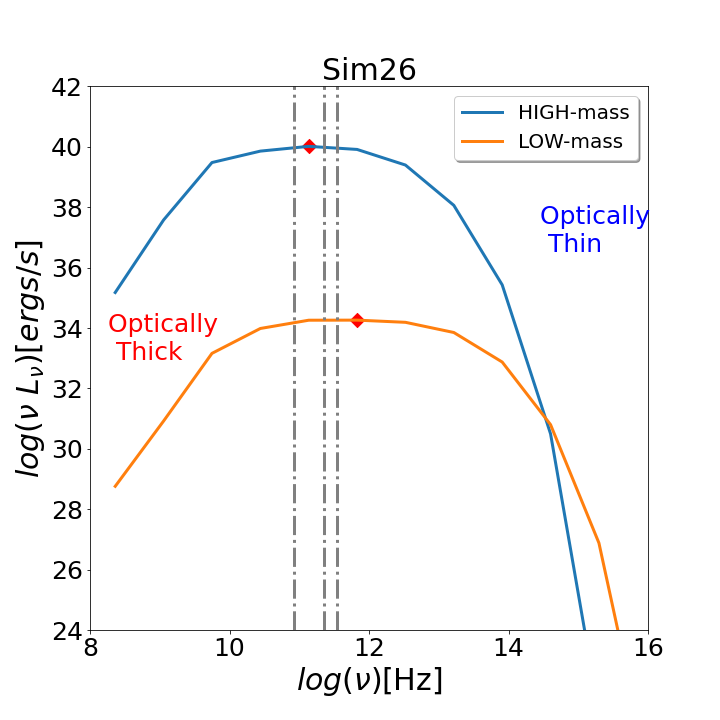}
  \caption{SEDs applying Eddington ratios from Table \ref{tab:para_dynamics} for a HIGH-mass (blue curves) and LOW-mass (orange curves) system for SIM20, SIM21, SIM22, SIM23, SIM24, and SIM26 respectively. The three vertical lines correspond to 86 GHz, 230 GHz, and 345 GHz, respectively. The figures also mark the optically thin and optically thick regions of the thermal synchrotron, which lie on the right and left sides of the thermal synchrotron peak marked by the red diamond.}
   \label{fig:specdiffM87SgrA}
\end{center}
\end{figure*}

\begin{figure}
 \begin{center}
 \includegraphics[height=2.8in]{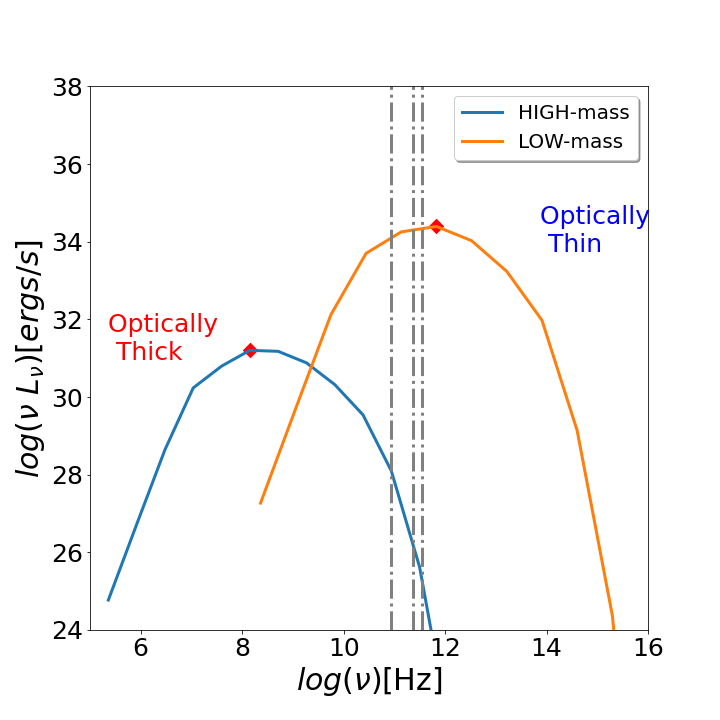}
   \caption{SEDs for simulations with black hole masses similar to M87 (blue curves, HIGH-mass) and SgrA* (orange curves, LOW-mass) for SIM20 with the same accretion rate. The three vertical lines correspond to 86 GHz, 230 GHz, and 345 GHz, respectively.
   The figure also marks the optically thin and optically thicker regions of the thermal synchrotron which lie on the right and left sides of the thermal synchrotron peak marked by the red diamond.}
   \label{fig:sameaccdiffM87SgrA}
\end{center}
\end{figure}

\begin{figure*}
 \begin{center}
 \includegraphics[height=3.5in]{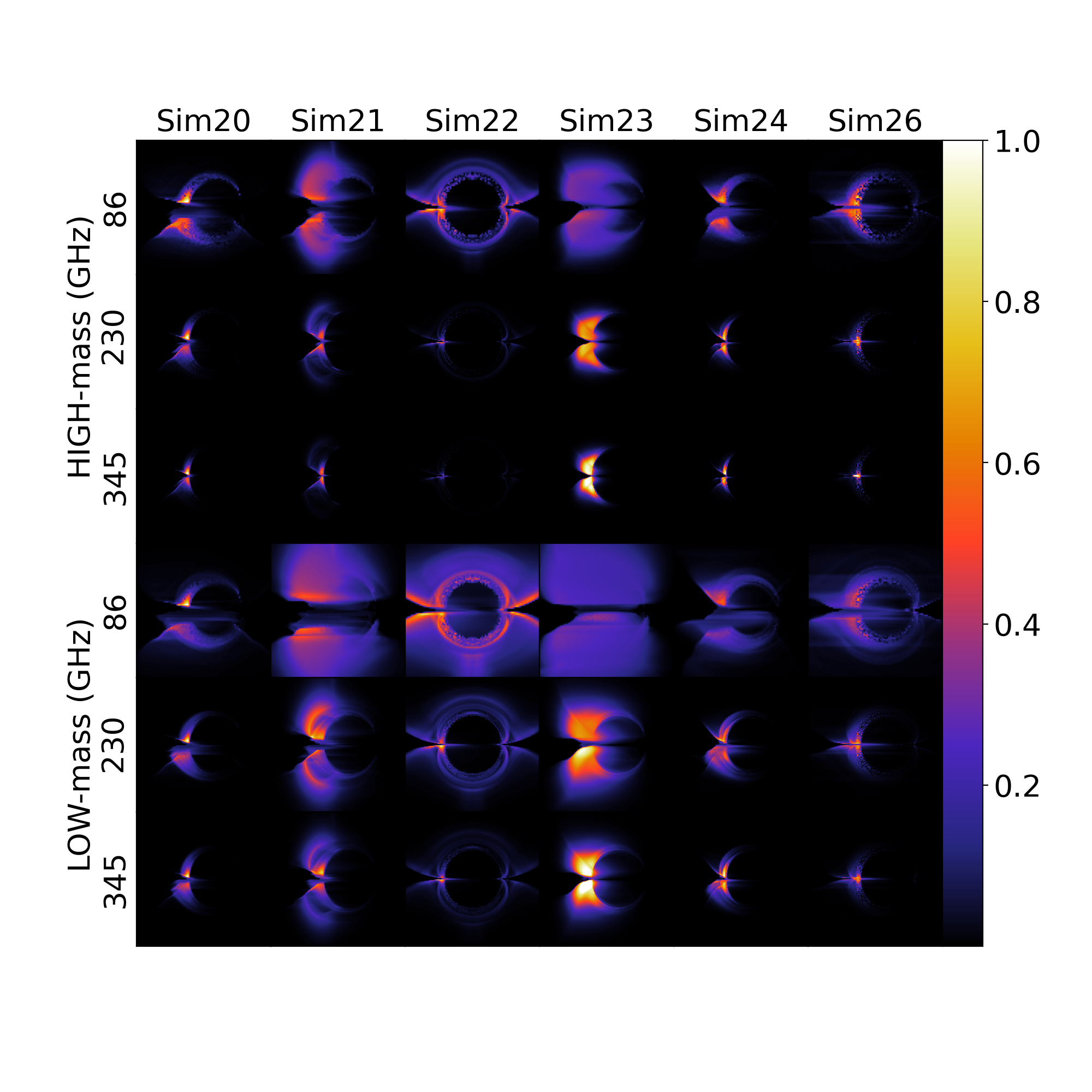}
  \includegraphics[height=3.5in]{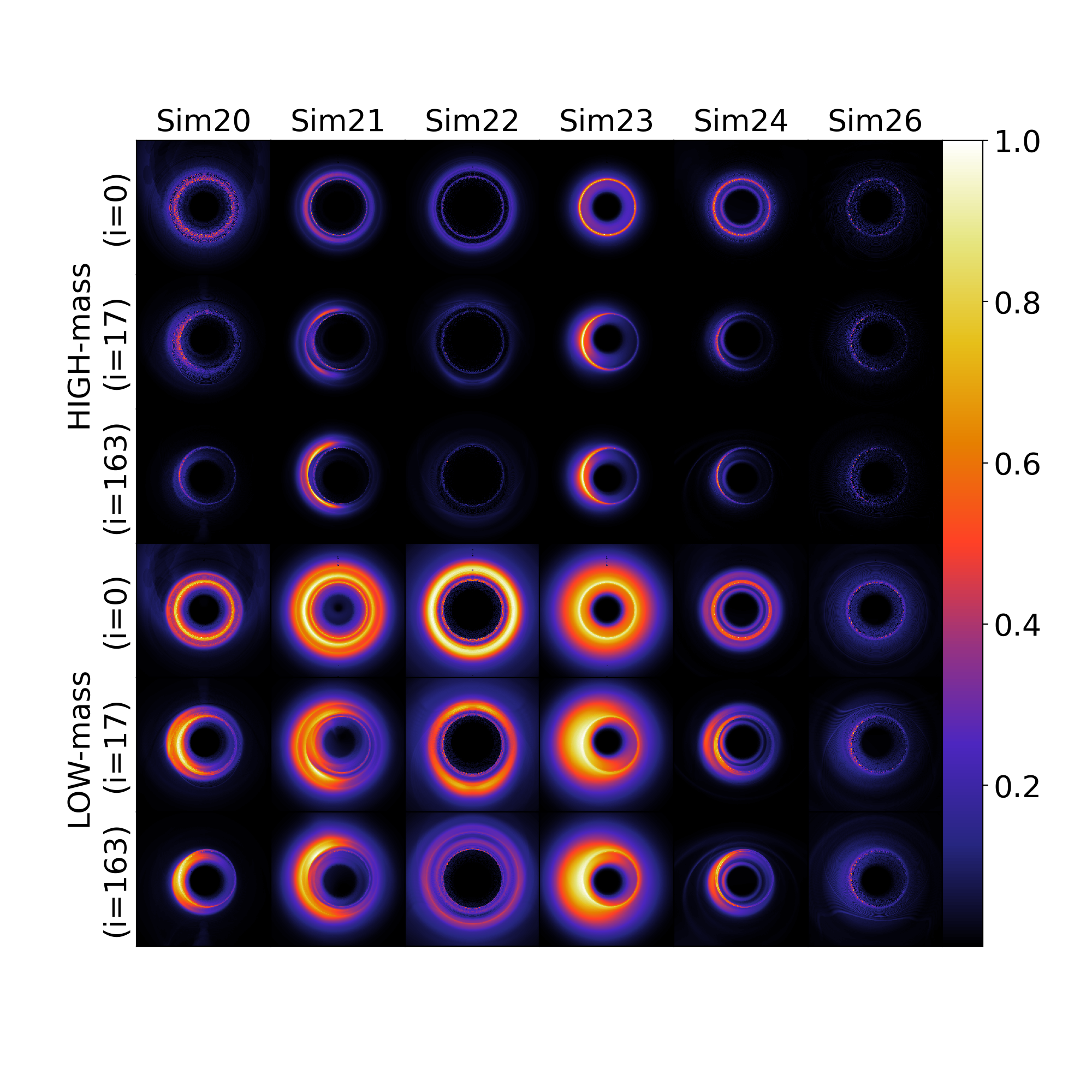}
   \caption{{\it Left:} Normalized emission maps (25 $R_g$ across) for an edge-on view for all our simulations applied for HIGH-mass (upper 3 rows) and LOW-mass (lower 3 rows) systems 
   at 86 GHz, 230 GHz and 345 GHz obtained with the Eddington ratio detailed in Table \ref{tab:para_dynamics}. {\it Right:} Normalized 230 GHz emission maps (25 $R_g$ across) as above but for inclination angles of $0^{\circ}$, $17^{\circ}$, and $163^{\circ}$.}
   \label{fig:allsimalli}
\end{center}
\end{figure*}

\begin{figure*}
 \begin{center}
 \includegraphics[height=3.5in]{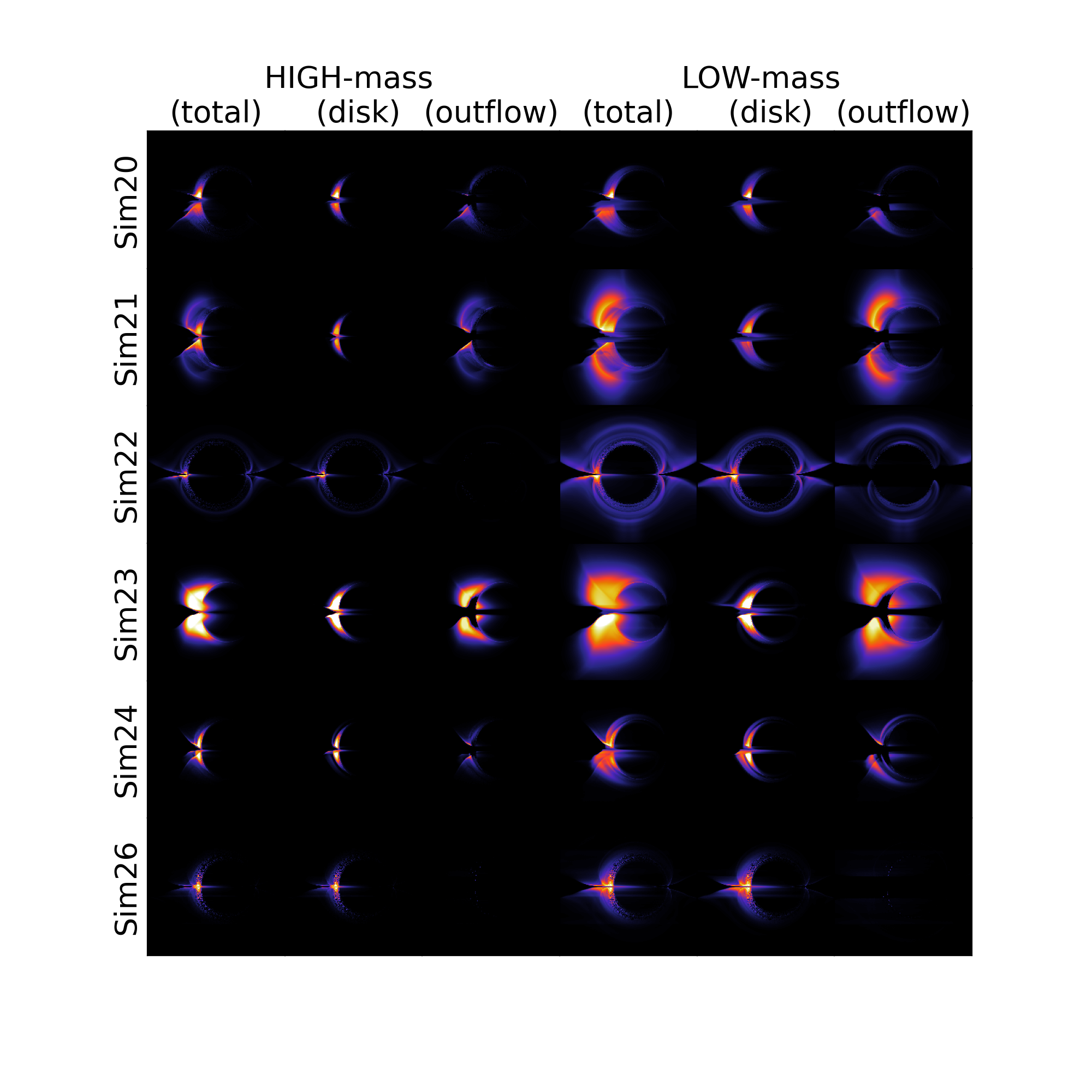}
 \includegraphics[height=3.5in]{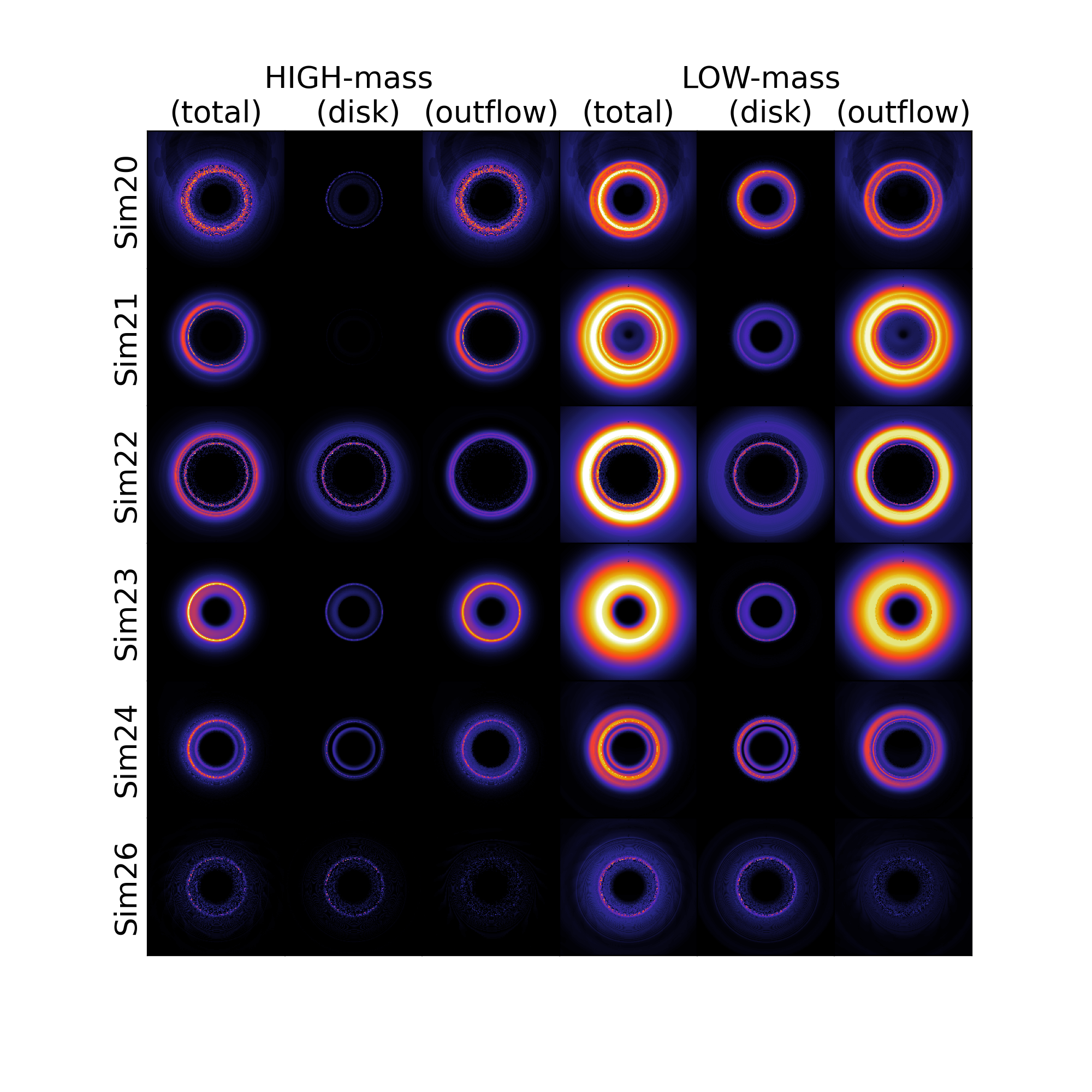}
   \caption{Normalized 230 GHz emission maps (25 $R_g$ across) for total, disk and "outflow" region normalized with the maximum of flux in the total emission map for edge-on inclination ({\it left}) and face-on inclination ({\it right}) for HIGH-mass (left 3 columns) and LOW-mass (right 3 columns) systems.}\label{fig:allsimallidiskoutflow}
\end{center}
\end{figure*}

\begin{figure}
 \begin{center}
  \includegraphics[height=3.5in]{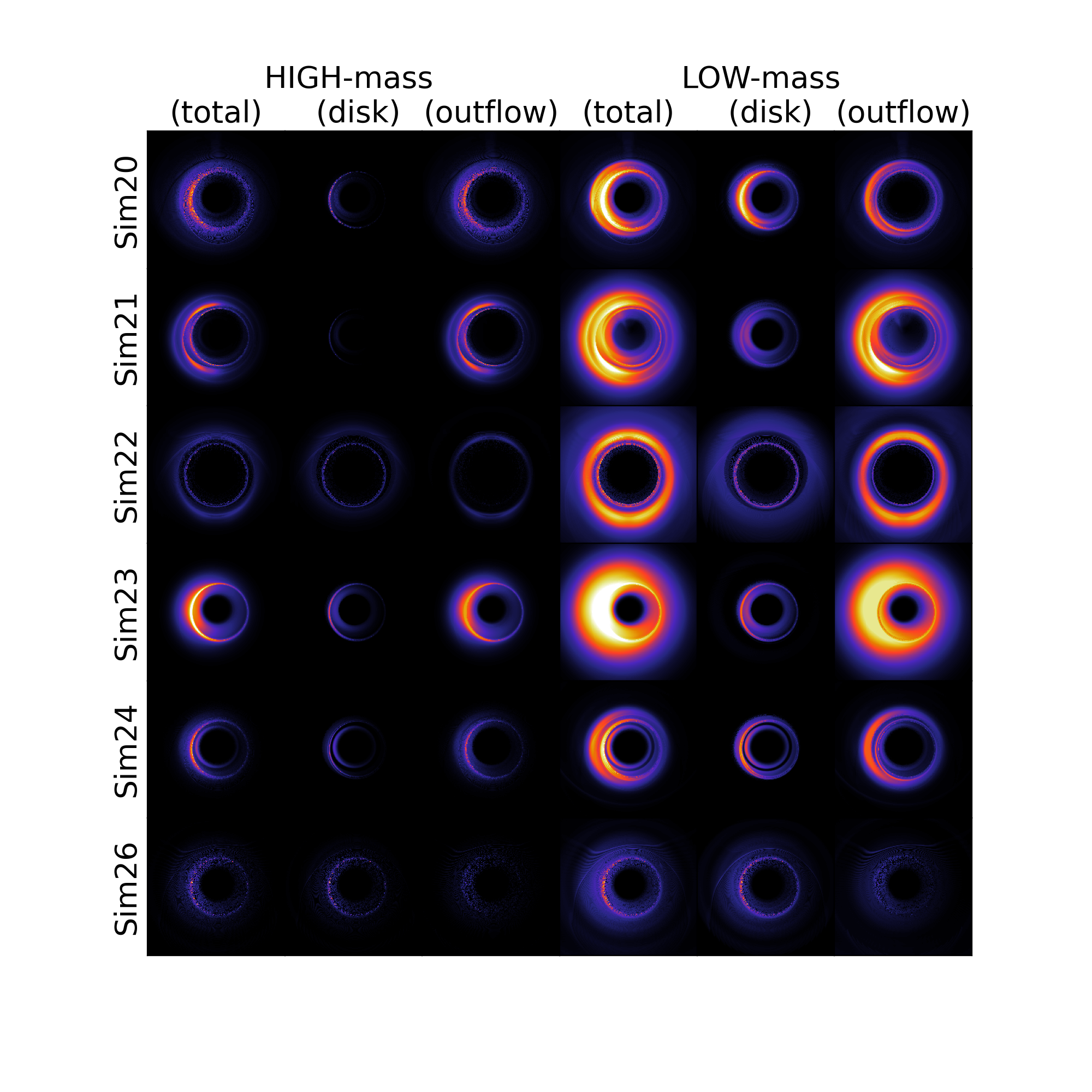}
  \caption{Normalized 230 GHz emission maps (25 $R_g$ across) for all the models with the total emissions, only the disk emissions and the emission from the "outflow" region for HIGH-mass (left 3 columns) and LOW-mass (right 3 columns) systems for an inclination angle of $17^{\circ}$.}\label{fig:prothdiskoutflowalli1}
\end{center}
\end{figure}

\begin{figure*}
 \begin{center}  
 \includegraphics[height=6.5in]{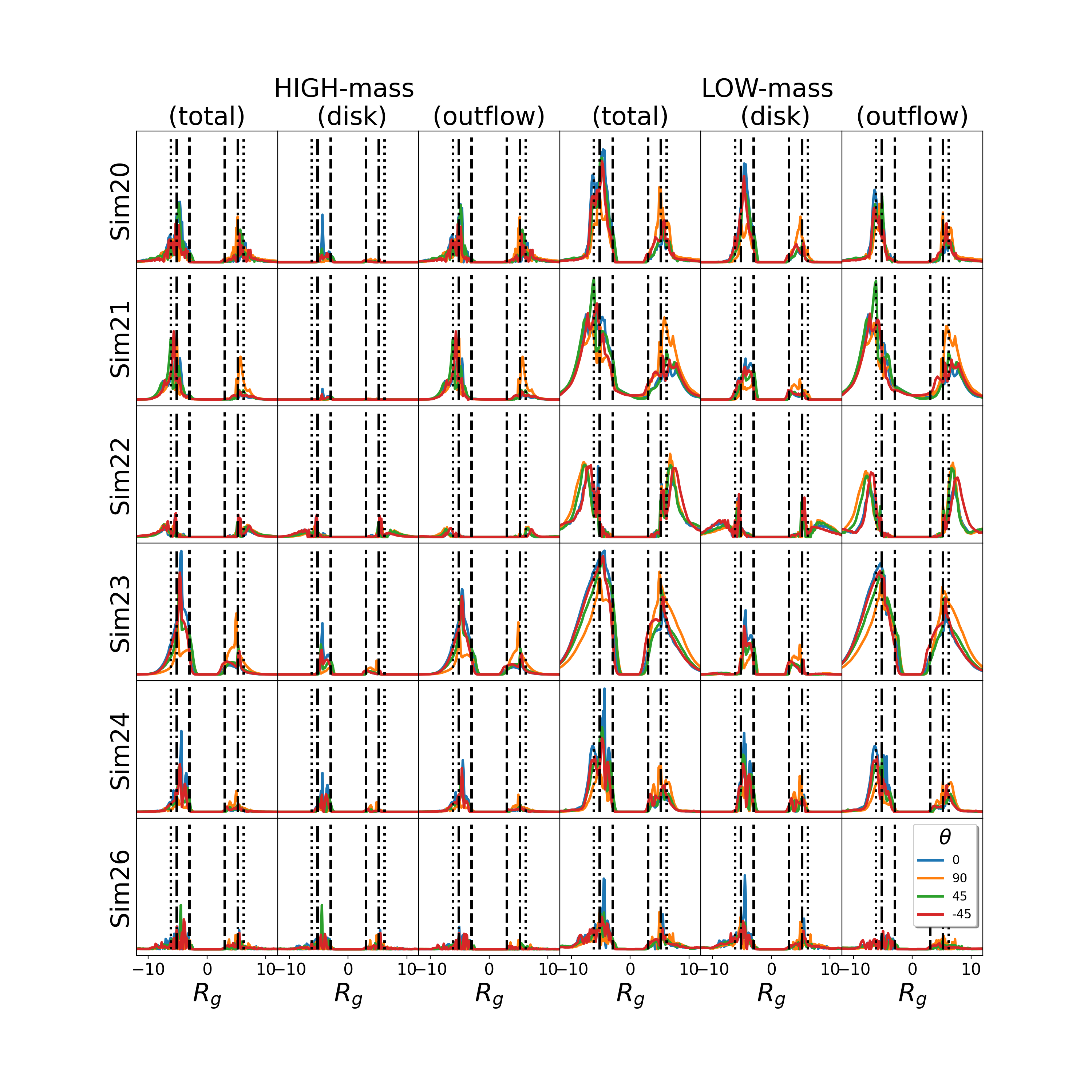}
  \caption{Normalized intensity profiles for the emission maps in Fig. \ref{fig:prothdiskoutflowalli1} along horizontal ($\theta=0\degr$, blue), vertical ($\theta=90\degr$, orange), and the two diagonals ($\theta=45\degr$, green and $\theta=-45\degr$, red) of the intensity map for all our simulations for a HIGH-mass (left 3 columns) and a LOW-mass (right 3 columns) system with an inclination angle of $17\degr$ for the total, disk and "outflow" regions. The intensity profile marked with a black solid line, a dot dashed line and a dashed line denote the positions of the unlensed photon-ring (3 $R_g$ ) and the outer and inner boundaries for the lensed photon ring (5.2 and 6.2 $R_g$, respectively).}\label{fig:prothdiskoutflowalli2}
\end{center}
\end{figure*}

\begin{figure}
 \centering
 \includegraphics[height=3.0in,bb=20 20 670 670]{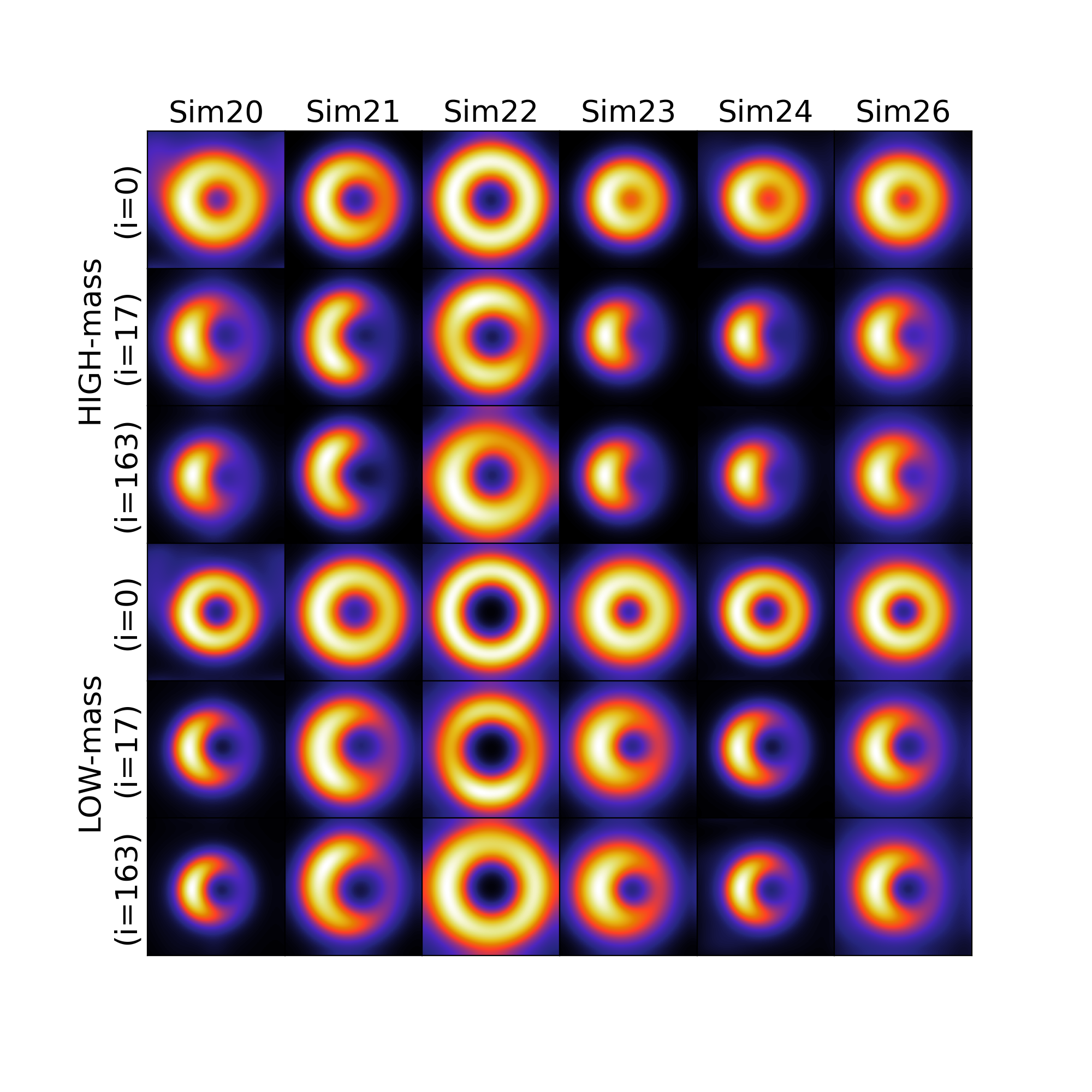}
 \includegraphics[height=3.0in,bb=20 20 670 670]{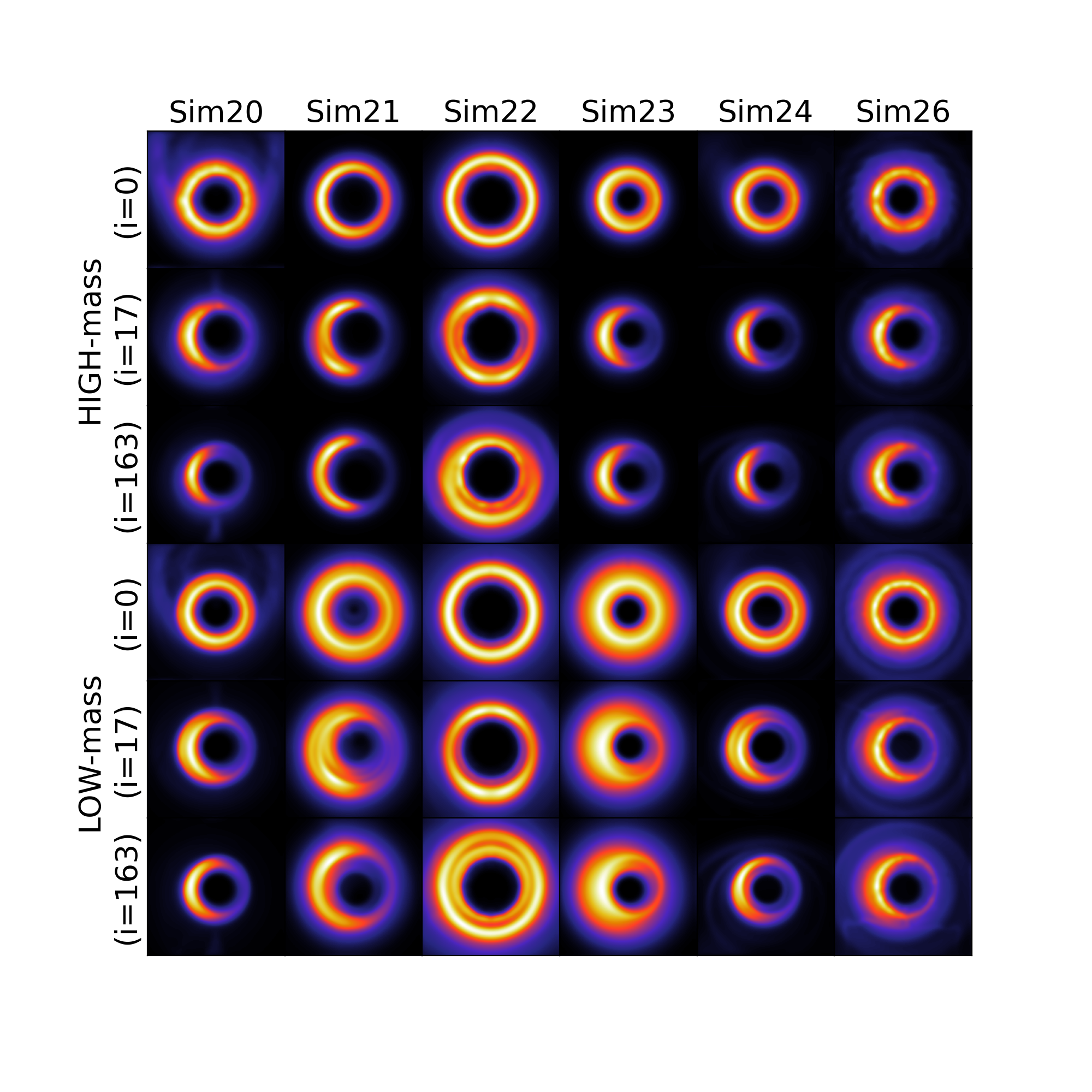}
 \includegraphics[height=3.0in,bb=20 20 670 670]{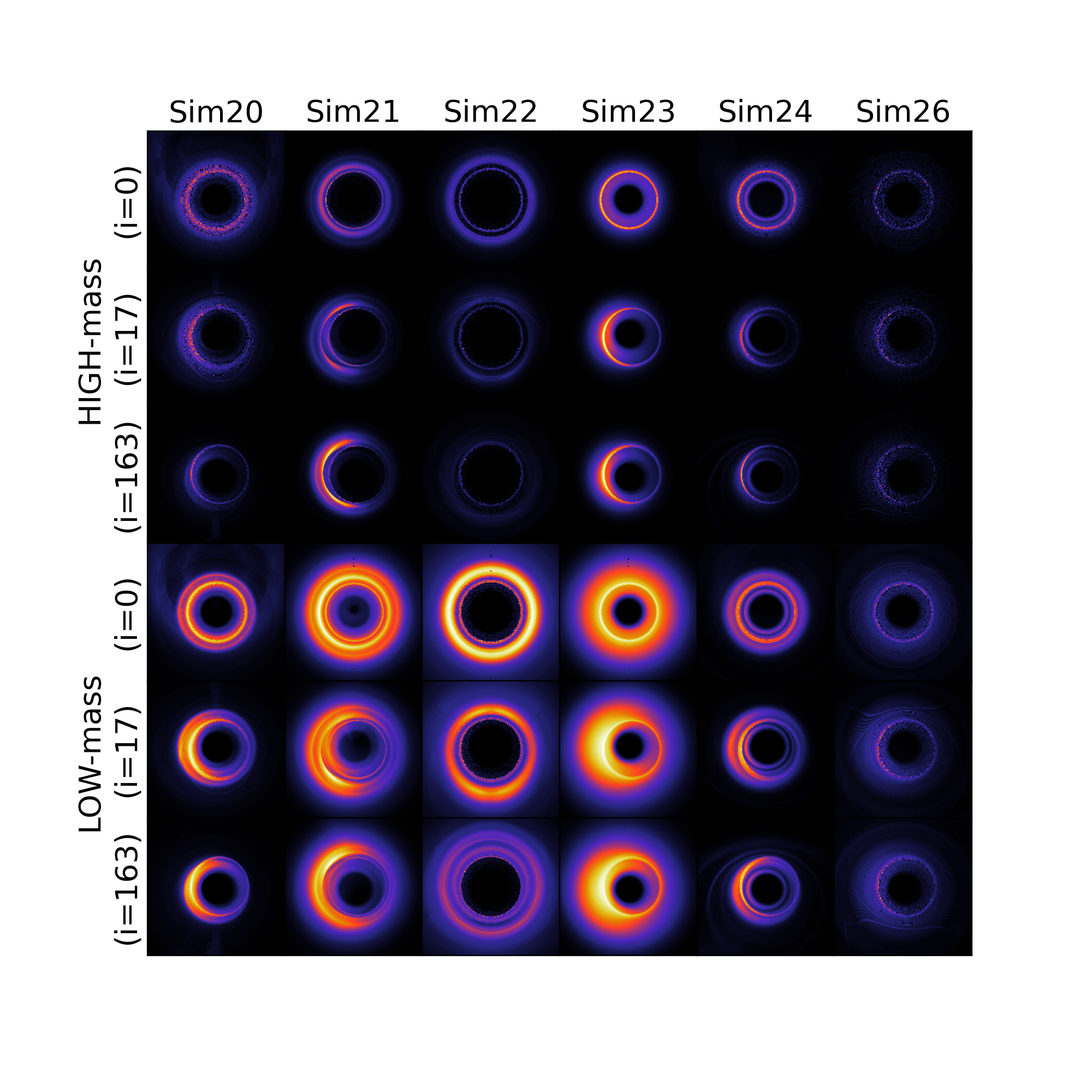} 
  \caption{Normalized blurred emission maps {(25 $R_g$ across)} for inclination angles of
  $0^{\circ}$, $17^{\circ}$, and $163^{\circ}$ for HIGH-mass (upper 3 rows) and LOW-mass (lower 3 rows) 
  systems assuming the resolutions with baselines that of EHT with $\theta=20~\mu as$ (upper left), 
  Geo-VLBI with $\theta=5~\mu as$ (upper-right) and L2-VLBI with $\theta=0.16~\mu as$ (bottom).} 
  \label{fig:allsimi17blurred}
\end{figure}

\section{Results: Radiation signatures} \label{sec:Results}
In the following subsections, we present our results in terms of the SED fitting, differences in the SED for the two mass models,  synthetic images,  and contribution to emission from the accretion and outflow regions, along with an analysis of the resolved synthetic images with different telescope resolutions. 

\subsection{Spectral fitting and analysis} \label{sub:Spec}
As mentioned in Sect. \ref{subsec:raytr}, the output of six GR-MHD simulations was postprocessed assuming masses similar to M87 (HIGH-mass) or SgrA* (LOW-mass). 
Taking the observed spectral data \citep{Prieto2016, Narayan1998} for these two systems as a reference, the Eddington ratio (and, thus, the accretion rate) was varied for each of these simulations, keeping all other parameters constant, to obtain a reasonable fit to the observed SED for each of these simulated models. 

The thermal synchrotron spectra generated thus for HIGH-mass and LOW-mass systems is shown in Fig. \ref{fig:spectrathM87SgrA}. The Eddington ratios thus obtained are provided in columns 7 and 8 of Table \ref{tab:para_dynamics}. A general comparison between a HIGH-mass and a LOW-mass system for all the simulations shows that the Eddington ratios obtained to fit the observed data of M87 are almost an order of magnitude higher than that required to fit to the observed SED of SgrA*. Thus, the total accretion rate required to obtain a spectral fit for a HIGH-mass system is higher -- not only because it is more massive, but also because it requires a larger reservoir of gas to be fed compared to a LOW-mass system such as SgrA*. This deduction is in agreement with what is generally inferred for M87 and SgrA*.

For a given simulation model, the low-frequency slope of the synchrotron spectrum was fixed, and the peak was shifted towards lower frequencies with increasing black hole mass. Once the lower-frequency end of the spectra has been determined from the underlying physical system and black hole mass, the spectral peak shifts towards higher (lower) luminosity with an increase (decrease) in the Eddington ratio or the accretion rate (see Fig. 6 of \citealt{Bandyopadhyay2021}). Thus, the spectral fits obtained here with our models are the best spectral fits. If the observed data lie below the spectral fits, those models are non-physical for those observed sources. Thus, in Fig. \ref{fig:spectrathM87SgrA}, models such as SIM26 and SIM20 can be completely ruled out for M87 SED, while SIM26 and SIM23 can be completely ruled out for SgrA* SED. 
However, since we intend to investigate the general properties of a HIGH-mass and LOW-mass system, all these model fits are assumed to be reasonable for this investigation.

In general, it was also observed that the Eddington ratios for systems with higher intrinsic magnetic field strength (lower $\beta_0$), such as SIM24 and SIM26 are lower, which is in accordance with the fact that the emissivity of the thermal synchrotron depends not only on the temperature, but also on the magnetic field strength. Simulations generated with higher density floor values (SIM21 and SIM23) require lower values of the Eddington ratio to generate the same fit, compared to those with lower floor values, such as SIM20. These high floor-valued simulations, although they may not physically exist, are investigated here solely for the purpose of understanding the emission features generated from the outflows in these systems. The Eddington ratio obtained for a spin zero case is the highest, which is expected, given that the ISCO radius is the largest for a non-spinning BH compared to a BH with prograde spin. 

\subsection{Synthetic spectra for HIGH-mass and LOW-mass source models}
The SED fits with the parameters of the HIGH-mass and LOW-mass sources are plotted over one another for each of the simulations in Fig. \ref{fig:specdiffM87SgrA}
 for a comparative study. The "red diamond" marker on these plots marks the maxima for each of these SEDs. 

A pure synchrotron SED is characterized by a low-frequency (below the peak-frequency) power-law slope resulting from synchrotron self-absorption, while the high-frequency slope is determined by the distribution function of the emitting electrons. This special property of a synchrotron emitting source makes it optically thicker at lower frequencies but optically thinner at higher frequencies.  

We observe in Fig. \ref{fig:specdiffM87SgrA} that while $230$ GHz and $345$ GHz lie on the high frequency slope of the SED for the HIGH-mass system, all frequencies of our interest lie on the self-absorbed slope for the LOW-mass system. This inference is in accordance with what is in general observed in AGN spectra with different masses. In \citet{Bandyopadhyay2021}, it was shown that for a given Eddington ratio, not only the peak luminosity increased with the BH mass but also the peak shifted to lower frequencies. Interestingly, if the accretion rates (not the Eddington ratios) are fixed, the peak frequency still decreases for a BH with higher mass, as shown in Fig. \ref{fig:sameaccdiffM87SgrA}. Thus, the location of the frequency (of observation or imaging) on the synchrotron spectrum results in interesting features in the emissions maps, as discussed in the next section. It has also been observed that for the simulations with higher intrinsic magnetic fields (i.e., SIM24 and SIM26), the peak is flatter (see Fig. \ref{fig:specdiffM87SgrA}) where the frequencies of interest in this investigation lie.

It is important to note that although the inferred Eddington ratios for each of these models are different, the frequency corresponding to the peak is the same regardless of the model. Thus, for these models with the given SEDs, the peak frequency is thus only a function of the mass of the BH. The peak frequency following \citet{Marscher1987} is $\nu_{peak} \propto B^{1/5}f^{2/5}(\nu_p)\Phi^{-4/5}$, where $f(\nu_p)$ is the peak flux and $\Phi$ is the field of view. As the peak frequency and field of view are constant for all the models, the small differences in the peak flux across the simulations are compensated for by the difference in the total magnetic field strengths across the simulations.

\subsection{Synthetic images} \label{sub:Intmap}
Once the Eddington ratios were obtained from the spectral fits to the data, ray-traced intensity maps were generated for a range of frequencies and inclination angles to understand the emission properties for each of these simulations. In general, it is observed that the emission regions for LOW-mass cases display extended bright regions in comparison to those for HIGH-mass cases. The brightest emission for the HIGH-mass systems is from the lensed photon ring. This difference occurs (as already mentioned in Sect. \ref{sub:Spec}) due to the special characteristics of the thermal synchrotron spectra (i.e., synchrotron self-absorption). In general, for sub-Eddington accretions at the frequencies we are interested in here, the LOW-mass AGN tends to lie on the optically thicker end of the thermal synchrotron spectra, as compared with those with higher masses.

The left panel of Fig. \ref{fig:allsimalli} demonstrates the edge-on views for the simulations for HIGH-mass and LOW-mass systems at 86 GHz, 230 GHz, and 345 GHz, respectively. In this figure, we can see that the regions closest to the BH are probed better with increasing frequencies. However, emission from the outer regions of the disk, outflows, or jets can be observed better at lower frequencies, such as 86-GHz emission maps for both HIGH-mass and LOW-mass systems, as they become optically thicker at lower frequencies because of synchrotron self-absorption. It is also observed that extended emissions at 86 GHz are visible for SIM21, SIM22, and SIM23 are possibly due to outflows from the disk. In addition, the Doppler beaming observed in the extended structure in SIM23 could arise due to the additional outflow velocity from the disk winds. The extended bright regions in SIM21 and SIM23 can also be seen at higher frequencies; however with decreasing emission brightness from outer regions.  

Outflows arising from regions in the vicinity of the BH may be easily identified for systems with edge-on view if the thermal synchrotron emission from these outflows are bright enough. However, in nature, many of the AGNs are found with different inclination angles. The inclination angle limits for both M87 and SgrA* suggest low inclination angles. The right of Fig. \ref{fig:allsimalli}, shows the emission maps at $230$ GHz for inclination angles of $i\approx 0^{\circ}$ and low inclination angles of $17^{\circ}$ and $163^{\circ}$ of the BH spin axis with respect to the line of sight. 

All of these simulations have the accretion flow rotating counter-clockwise with respect to the BH spin axis, which is also validated from looking at the edge-on emission maps in Fig. \ref{fig:allsimalli} where the west of the images are brighter due to Doppler beaming effects in counter-clockwise rotating systems. Doppler beaming for such anti-clockwise rotations would thus lead to an increased brightness south-westward for inclination angles between $0-90$ degrees and north-westward for inclination angles between $90-180$ degrees. All simulations except SIM22 and SIM23, are brightest south-westward and north-westward for inclination angles of $17^{\circ}$ and $163^{\circ}$ respectively. In SIM22,  the north and south regions are the brightest for $17^{\circ}$, however, for SIM23, the brightest regions are opposite to what would be physically expected. This could be due to additional velocity field effects arising from "outflow" regions.

\subsection{Emission from disks and "outflow" regions}
To understand the contribution to the emission from the disk and the "outflow" regions, 
we should consider the emission maps from these regions separately.
This is shown in Fig. \ref{fig:allsimallidiskoutflow} for edge-on and face-on inclinations and Fig. \ref{fig:prothdiskoutflowalli1} for low inclination angle views, as described in Sect. \ref{subsec:raytr}. 
The figures show the emission from the total, disk, and "outflow" regions separately. 

In the edge-on case, the disk emissions from all the simulations, except for SIM22 and SIM23,
appear to be similar. 
The dominant contribution to the total emission appears to be from the disk, except for simulations SIM21 and SIM23. 
However, for all low-inclination angles, the maximum contribution to the total emission is from
the "outflow" region for all simulations except for SIM26.
When the projection of the jet axis along the line of sight is not zero, there is an enhancement in brightness due to Doppler beaming. 

The differences in emissions from the disk and "outflow" regions are better explained with the emission profiles for the $17^\circ$ emission maps shown in Fig. \ref{fig:prothdiskoutflowalli2}. These emission profiles are obtained by cutting slices along the x-axis (blue line), y-axis (orange line), and the two diagonal lines (green line and red line) from the emission maps. It can be inferred from these emission profiles that in general, for an inclination angle of $17^{\circ}$, the emission is dominated by the "outflow" region, except for SIM26. 

Disk emission has a maximum along the x-axis slice (blue line) except for SIM22. The peak brightness in these cases lies close to $5.2~R_g$ and, in general, disk emission is restricted within $6.2~R_g$, except for SIM22. 
However, the outflow emissions extend wider and the maxima can be located anywhere between $-45^{\circ}$ and $45^{\circ}$.
Only for SIM26, the total emission is dominated by the disk emission, as also observed for the other inclination angles.

We also observe that for black holes with non-zero spin, there is a peak asymmetry, with the left peaks higher than the right except for SIM22. 
Simulation SIM22 has both peaks at an equal height as is expected for a Schwarzschild BH. 
In general, it is also observed from the emission profiles that they are broader and noisier for the LOW-mass models than for the HIGH-mass models, since synchrotron self-absorption has more of an effect on a LOW-mass system than on a HIGH-mass system, as explained above. 
This effect is generally more pronounced in the "outflow" regions. 
In optically thicker systems, low density "outflow" regions brighten up and are noisier, compared to the disk emission.

\subsection{Resolved images with different telescope resolutions applied}
The emission maps seen in Fig. \ref{fig:allsimalli} are idealized images generated through ray-tracing procedures; however, in reality, observations are limited by telescope resolution, which greatly depends on the telescope diameter. At present, observations with the highest resolutions are possible with an Earth-sized diameter, made possible by the EHTC. However, to improve upon the resolution further, it would be necessary to resort to space-VLBI facilities. In Fig. \ref{fig:allsimi17blurred} idealized resolved images are shown, where the emission maps for those on the right of Fig. \ref{fig:allsimalli} have been convolved with a Gaussian beam-width for three telescope resolutions: EHT ($20~\mu$as-upper left block), Geo-VLBI-like baseline ($5~\mu$as- upper right block), and L2-VLBI-like baseline ($0.16~\mu$as- lower block) at 230 GHz. With the EHT-like resolution, some of these models appear degenerate, such as SIM23 and SIM24 as in HIGH-mass systems.

The difference in the extent of the emission features such as the ones seen for SIM21 and SIM23 for the LOW-mass system is quite degenerate at current resolutions. However, as we have already seen, with an EHT-like resolution, it could be easy to distinguish a spinning BH from a non-spinning one, as it can be seen from the clear asymmetrical features as well as the smaller ring diameter observed for all the simulations with non-zero spin systems (SIM20, SIM21, SIM23, SIM24, and SIM26) than seen for the spin zero case (SIM22). 

It is also interesting to note here that the difference in emission features that was observed for the HIGH-mass and LOW-mass systems in Fig. \ref{fig:allsimalli} is smoothed out by the limiting telescope resolution.  The contrast in the image features improves with Geo-VLBI-like resolution, however, they are still quite degenerate, especially for HIGH-mass systems.

\section{Discussion}\label{sec:Discussion}
Overall, SMBHs are expected to be found in the centers of galaxies, as well as in the heart of other AGNs. However, a multi-wavelength observation of such systems can provide important information related to the inflow and outflow mechanism around each of these systems. Some of these sources may have powerful jets, while others may be in a more quiescent state. Now with the two years of observations for M87 by the EHTC \citep{EHTC2019a,EHTC2024a} and the first image of SgrA* \citep{EHTC2022a}, as well as the polarized imaging of them \citep{EHTC2021, EHTC2024b}, it has been possible to probe regions in the vicinity of these supermassive objects. The difference in these images piqued our curiosity about the range of physical processes that led to these interesting observations. 

In this work, we used a set of six models, which were simulated with the assumption of a resistive thin Keplerian disk under different initial conditions, evolved over time, and led to different accretion and outflow structures in their final stable configuration. As with all GR-MHD simulations, these simulations are in code units and need to be postprocessed with different physical parameters to provide an understanding of a real physical system. Here, we have chosen the masses and spectral data of M87 and SgrA*  \citep{Prieto2016, Narayan1998} to distinguish our models and the various emission characteristics arising from these models under these constraints. Thus, the goal of this work is not to provide alternative models for M87 and SgrA*, but instead to use certain characteristics of these systems to analyse the properties and differences between the models through their emission features.

The primary goal of this investigation has been to understand the emission characteristics that can arise from different physical systems using a range of models that have intrinsically various structures, such as outflows from disk winds and jets, as described in detail in \cite{Bandyopadhyay2021} with a given mass and SED. We also explore the possibility of understanding whether these features can be distinguished with the current telescope resolutions or, as an alternative, the baselines that need to be used to extract such features. 
As mentioned, the SED and masses of M87 and SgrA* serve as reference systems, as both these sources have not only been extensively investigated; in addition, their first images act as a basis for studying a HIGH-mass and LOW-mass system. 
Our primary deductions from this investigation can be summarized as follows.
\begin{itemize}
    \item The peak frequency of the SED for the models depends inversely on the BH masses through the magnetic field strength. However, the shape of the SED depends on the underlying model.
    \item The models are optically thinner at 230 GHz and 345 GHz under HIGH-mass constraints where the emission is dominated by the underlying distribution of the thermal electrons; whereas for the frequencies of interest here, LOW-mass systems are optically thicker, as inferred from their thermal synchrotron SED where the emission is dominated by synchrotron self-absorption.  
    \item In cases of optically thicker systems (as in LOW-mass constraints),  emission from the outflow regions becomes more prominent. However, for an optically thinner case (as at 230 GHz for HIGH-mass constraint), the emissions from the outflow regions are too low compared to the maximum emission from the lensed photon ring. 
    \item For most models, the maximum contribution to the total emission is from the "outflow" regions, except the one with the highest intrinsic magnetic field strength.
    \item Although the disk emission features are quite similar for all the Kerr BH models, the emission from all the models significantly differs in brightness features, depending on the viewing angles. A significant enhancement in brightness at low viewing angles in models with strong outflows is brought about by Doppler beaming.
    \item With the $20~\mu as$ resolution of the EHT, it is not possible to distinguish one model from the other with the intensity maps (Stokes I) alone. However, the differences in these model emissions can be observed with higher resolutions such as an L2(Space)-VLBI baseline.
\end{itemize}

With this exploratory investigation, we observe that even though the total luminosity for HIGH-mass models is higher with the inferred accretion rates also higher compared to the LOW-mass models, they appear to be optically thinner due to the special characteristics of the synchrotron SED, where the synchrotron self-absorption band extends to higher frequencies for lower mass black holes than massive black holes. We have shown that a massive system appears to be optically thinner even for similar accretion rates. This is an important characteristic of a massive system like M87 for which it might be easier to extract or test characteristics related to the photon ring such as different GR models; however, a detection of models related to outflows or inflows close to the event horizon can be better understood with a less massive system with similar underlying models.

We also demonstrate in this work that for most of the models except the one with the highest magnetic field, the maximum contribution to the total emissions is from the "outflow" region and not from the pure accretion inflow region. It is inferred that the emission brightness changes significantly for the "outflow" regions for different inclination angles. As we mentioned previously and also shown in \citet{Bandyopadhyay2021} that some of these GR-MHD models have large-scale jets; however, the strength of these jet emission is a work in progress (Bandyopadhyay et al. in preparation) and is beyond the scope of this work.

Although the models utilized in the present work were simulated to launch large-.scale jets and outflows under different initial conditions, in the future, we primarily aim to explore the emission from these models close to the event horizon, as it has been of great importance to understand the jet launching mechanisms and underlying physics close to the event horizon. The flux properties from the large-scale outflows from these models is being investigated (Bandyopadhyay et al. in preparation), and further constraints on these models can be provided with such an investigation. Thus, the primary significance of this work lies in separating out the emissions individually from the disk or "outflow" regions. 

It is important to note that while it might be more difficult to distinguish the emission from the accreting material and outflow close to the horizon in MAD/SANE models (as the density contrast between the accretion and outflow region is low), on the other hand, our models have higher density contrast between the accretion and "outflow" region, which makes the emission from "outflow" region quite prominent (in particular for LOW-mass systems) especially at low inclination angles. In addition, these resistive GR-MHD models enable the launching of strong disk jets; this makes the jets more easily visible at lower inclination angles due to Doppler beaming. 

Some of the models investigated here may not exist physically (due to certain assumptions such as high floor values); however, they have been studied to understand the emissions from the outflows in these systems. As deduced by the EHTC for M87 and now also for SgrA*, better constraints on the underlying physics are obtained by investigating the polarized emissions from them close to the event horizon \citep{EHTC2021,EHTC2024b}. However, such a detailed investigation on the polarized emissions from these models is left for future work.

The Eddington ratio (and thus the accretion rate) inferred from our investigation is similar to those inferred for M87 \citep{EHTC2019a} and SgrA* \citep{EHTC2022a} using MAD/SANE models, which in general are geometrically thicker than the Keplerian thinner disk in this investigation. This results in our models being optically thicker in general than those of the standard MAD/SANE models. Also, the SEDs for these systems as shown in the literature have peak frequencies which are higher than the 230 GHz and 345 GHz for both M87 and SgrA*, implying that at these frequencies both these sources lie on the synchrotron self-absorption slope of the synchrotron SED. However, for a thinner Keplerian disk model as has been shown in this work, although for LOW-mass these frequencies lie on the synchrotron self-absorption slope; however, for HIGH-mass, these frequencies lie on the higher frequency slope where the source appears to be optically thinner even though the accretion rates and, thus, the densities are higher for HIGH-mass systems. 

In addition, the peak frequency being the same for a given black hole mass irrespective of the simulated model implies that the peak frequency is determined by the emission from the shape of the accreting disk (shown here) and is similar for all our models. We can further add that any model for which the highest VLBI frequencies attainable lie below the peak frequency for HIGH-mass constraints will also be the same for a LOW-mass system, but the reverse is not true. However, even in those cases, a LOW-mass system will appear to be optically thicker than a HIGH-mass system.

The unresolved maps of the different models presented here show significant differences, but with an EHT like resolution of $20~\mu$as, the resolved maps appear to be similar to a greater extent with only a significant difference in these emission maps for the spinning or non-spinning BHs. However, with better resolutions with space VLBI, such as L2-VLBI baselines and improved imaging algorithms, the finer details in the images can be captured; thus, this approach can also give us a better physical insight about the processes. 

Resorting to space VLBI will thus not only help us to image features in M87 and SgrA* with greater detail, but also make it possible to image the BH shadows of objects with smaller angular ring diameters. Such resolutions can also reveal the existence of higher order rings ($n=1,2$ etc.). The ngEHT will already be an improvement with better image fidelity (more $u-v$ coverage) and sensitivity with a resolution gain from moving to 345 GHz, as well as super-resolution from imaging \citep{Chael2016,Akiyama2017} and Bayesian imaging and modelling techniques \citep{Broderick2020}. It is important to note that the blurring introduced here is only due to the limited telescope resolution. However, further blurring can be caused by insufficient u-v coverage. Thus, not only is it important to improve the baseline, but it is also important to strive for better u-v coverage. 

\section{Conclusions} \label{sec:Conclu}
In this work, we have shown how the underlying AGN inflow and outflow dynamical model affects the spectral shape and produces interesting emission features close to sub-mm wavelength bands. We used the data of resistive GR-MHD models, where the underlying accretion flow has a thinner Keplerian disk structure.
The underlying GR-MHD dynamics should be understood as a toy model that is able to provide distinct dynamical features,
such as strong disk winds in combination with a fast, black hole driven spine jet, in addition to the disk structure and the black hole magnetosphere.

In particular, we have shown that given a model, the emission properties at a particular frequency are completely determined by where the frequency lies on the synchrotron SED, which is determined by the mass and accretion rate of the system concerned. 
For low-mass and lower Eddington ratios, most of our models display strong emission features from disk winds and jets at 230 GHz, while the same features display primarily the ring emission for high-mass systems. 

In general, for lower frequencies on the self-absorption synchrotron slope, the system is optically thicker; thus, with lower frequencies, the low-density regions (e.g., as outflows) become brighter. 
The decrease in emission at frequencies greater than the peak frequency is quite stark as the system becomes optically thinner and the emission follows the distribution of the underlying electrons. Thus, to be able to probe the emissions from outflows on close to horizon scales, it is important for the emission spectrum to be at a level where the peak frequency of the emission spectrum is greater than the frequency of interest. 

Concerning observations, the degree of freedom of observing the system at different inclination angles will add to other observational degeneracies. 
It would thus be important to investigate these systems at multiple frequencies along with other observational probes such as polarization, multi-year analysis, etc., to trace the emission signatures from outflows. 
The Eddington ratios we infer from the SED fitting of thin Keplerian disk models are similar to those deduced by the EHT for M87 and SgrA* with MAD/SANE models;
however, in those models, the accretion flow is geometrically thicker. 
A similarity in the accretion flows for such differences in geometry implies that a disk model must be denser and thus optically thicker, respectively. 
After all, these differences in the disk geometry can, along with other factors, also contribute to the variation we observe in the resulting shape
of our spectra.

Observations of M87 and SgrA* have already shown that these systems are really important not only because of the very different black hole masses but also because of various different physical properties. An example is the recent M87 image with 2018 data, which has shown that the brightest emission region is rotated by almost 30 degrees \citep{EHTC2024a} from the previous image, obtained in 2017 \citep{EHTC2019a}. 

These asymmetries can arise due to variation in the accretion inflow or perhaps due to some outflow, which can be inferred in due course by observing the systems over years or in frequent time intervals, including observations of polarized signals. 
As the models investigated here are axisymmetric, we do not expect to observe such variation from the accretion flows for such timescales. However, in axisymmetric models, sporadic flows from the disk winds may result in changes in the brightest emission region over time.

A fully 3D dynamical treatment would bring in further features that cannot be traced in a 3D-axisymmetric modeling as we did. 
This will include structural changes, such as localized hot spots and different time scales of variability.
More elaborate methods will be needed to fully disentangle the properties of inflows from those of outflows. 
Ideally, a slow-light approach would be needed to ray-trace relativistically moving substructures realistically.

So far, with our approach, we have been able to identify that the current EHT resolution of 20 $\mu$as cannot fully distinguish the different physical models; therefore, a larger baseline (even as large as L2-VLBI) will be necessary to provide more detailed observations that could alleviate certain observational degeneracies. 
It is expected that future observations and data analyses of many such AGN systems will thus provide better prospects and improve our understanding of the accretion mechanism in them. 

\begin{acknowledgements}
We acknowledge funding from ANID Chile via Nucleo Milenio TITANs (NCN2023$-$002), Fondecyt 1221421, and Basal projects AFB-170002 and FB210003. 
B.B. thanks the CAS-ANID funding via project CAS220010.
We thank Javier Lagunas for valuable discussions and collaborations.
C.F. acknowledges the support of the German Research Foundation DFG via the Research Unit FOR\,5195. 
All GR-MHD simulations were performed on the VERA cluster of the Max Planck Institute for Astronomy. 
C.F. thanks Qian Qian and Christos Vourellis for developing the resistive version rHARM-3D of the original GR-MHD code HARM-3D, kindly provided by Scott Noble. FAS is supported by the Spanish FPI fellowship from AEI grant PID2023-152148NB-
I00.

\end{acknowledgements}

\bibliographystyle{aa}
\bibliography{M87SgrA}

\end{document}